\documentclass[prb,aps,super,
twocolumn,
print,%linenumbers,
groupedaddress,superscriptaddress,
amsfonts,amssymb,amsmath,floatfix,
showkeys,showpacs,a4paper]{revtex4-1}

\usepackage{graphicx}
\usepackage[english]{babel}
\usepackage{hyperref}
\usepackage{upgreek}
\usepackage{natbib}
\usepackage{lineno}
\usepackage{color}
\usepackage{multirow}
\usepackage[T1]{fontenc}       % Use modern font encodings
\usepackage{graphicx}
\usepackage[english]{babel}
\usepackage{hyperref}

\begin{document}

\title{Epitaxial strain adaptation in chemically disordered FeRh thin films}

\author{Ralf Witte}\email{ralf.witte@kit.edu}
%\affiliation{Institute of Nanotechnology, Karlsruhe Institute of Technology, 
% 76344 Eggenstein-Leopoldshafen, Germany}
\author{Robert Kruk}
\affiliation{Institute of Nanotechnology, Karlsruhe Institute of Technology, Hermann-von-Helmholtz-Platz 1,
 76344 Eggenstein-Leopoldshafen, Germany}
 \author{Di Wang}
 \affiliation{Institute of Nanotechnology, Karlsruhe Institute of Technology, Hermann-von-Helmholtz-Platz 1,  
 76344 Eggenstein-Leopoldshafen, Germany}
 \affiliation{Karlsruhe Nano Micro Facility (KNMF), Karlsruhe Institute of Technology, Hermann-von-Helmholtz-Platz 1, 76344 Eggenstein-Leopoldshafen, Germany}
\author{Sabine Schlabach}
  \affiliation{Karlsruhe Nano Micro Facility (KNMF), Karlsruhe Institute of Technology, Hermann-von-Helmholtz-Platz 1, 76344 Eggenstein-Leopoldshafen, Germany}
  \affiliation{Institute for Applied Materials, Karlsruhe Institute of Technology, Hermann-von-Helmholtz-Platz 1,  
 76344 Eggenstein-Leopoldshafen, Germany}
\author{Richard A. Brand}
\affiliation{Institute of Nanotechnology, Karlsruhe Institute of Technology, Hermann-von-Helmholtz-Platz 1, 
 76344 Eggenstein-Leopoldshafen, Germany}
\affiliation{Faculty of Physics and Center for Nanointegration Duisburg-Essen (CENIDE), University of Duisburg-Essen, 47048 Duisburg, Germany}
\author{Markus E. Gruner}
\affiliation{Faculty of Physics and Center for Nanointegration Duisburg-Essen (CENIDE), University of Duisburg-Essen, 47048 Duisburg, Germany}
\author{Heiko Wende}
\affiliation{Faculty of Physics and Center for Nanointegration Duisburg-Essen (CENIDE), University of Duisburg-Essen, 47048 Duisburg, Germany}
\author{Horst Hahn}
\affiliation{Institute of Nanotechnology, Karlsruhe Institute of Technology, Hermann-von-Helmholtz-Platz 1, 
 76344 Eggenstein-Leopoldshafen, Germany}
\affiliation{KIT-TUD-Joint Research Laboratory Nanomaterials, Technical University Darmstadt, Jovanka-Bontschits-Str. 2,  64287 Darmstadt, Germany}

\date{\today}

\begin{abstract}

Strain and strain adaptation mechanisms in modern functional materials are of crucial importance for their performance. Understanding these mechanisms will advance innovative approaches for material properties engineering.  Here we study  the strain adaptation mechanism in a thin film model system as function of epitaxial strain. Chemically disordered FeRh thin films are deposited on W-V buffer layers, which allow for large variation of the preset lattice constants, e.g. epitaxial boundary condition.
It is shown by means of high resolution X-ray reciprocal space maps and transmission electron microscopy that the system reacts with a tilting mechanism of the structural units in order to adapt to the lattice constants of the buffer layer. This response is explained by  density functional theory calculations, which evidence an   energetic minimum for structures with a distortion of $c/a\approx0.87$. The experimentally observed tilting mechanism is induced by this energy gain and allows the system to remain in the most favorable structure. In general, it is shown that the use of epitaxial model heterostructures consisting of alloy buffer layers of fully miscible elements and the functional material of interest allows to study strain adaptation behaviors in great detail. This approach makes even small secondary effects observable, such as the directional tilting of the structural domains identified in the present case study.

\end{abstract}

\maketitle

\section{Introduction}

Strain and stress at interfaces in-between different phases  are omnipresent in modern functional materials. These interface effects may lead either to a degradation or more precisely an undesired property modification due to the strain in the material or, the more fortunate case, by making use of strain engineering beneficial properties of the material may be designed. An example for the latter case is the epitaxial growth of oxide heterostructures \cite{Schlom2007a}, allowing for a delicate adjustment of functional ferroelectric properties with the choice of the single crystal substrate. 
In addition  it is well known that properties of semiconductor heterostructures such as charge carrier mobility, can be modified by orders of magnitude by using different substrates or buffer materials \cite{DelAlamo2011}. 

In addition to these examples from thin film technology, there are also cases in which  interfaces form within one material, which goes through a phase transition. 
A prominent case is given by martensitically transforming materials, where semi-epitaxial interfaces are formed between the martensite and austenite phases, which in this context are referred to as habit planes. Along these interfaces, large strains occur leading to special adaptation mechanisms, such as adaptive martensite structures \cite{Khachaturyan1991,Bhattacharya2004} in shape memory alloys (SMA) \cite{Kaufmann2010,Niemann2012}. Such martensitically transforming materials are interesting for medical or actuator applications. Magnetic SMA (MSMA) are  attractive materials for magnetomechanical actuators\cite{Ullakko1996,Sozinov2002} and magneto-caloric cooling \cite{Kainuma2006,Planes2009,Liu2012}. For these applications, the formation of the adaptive nanotwinned interfaces is of crucial importance for the functional properties and reversibility of the materials\cite{Gruner2018}. Hence the mechanisms of interface formation and the occurring strain adaptation behaviors are of great interest for the design of these complex functional materials.

The present study presents a systematic investigation of unprecedented strain adaptation behavior found in equiatomic Fe-Rh alloy thin films. The FeRh binary alloy system has attracted substantial research efforts, due to a peculiar magnetic transition, occurring in the equiatomic B2-ordered (CsCl structure) alloy. Above room temperature an antiferromagnetic-to-ferromagnetic transition was observed in the 1930s\cite{Fallot1938,Fallot1939}, which  still continues to motivate various research activities \cite{Wolloch2016,Lewis2016,Kim2016,Aschauer2016a,Bennett2018,Gorji2018,Drozdz2018,Popescu2018} and has triggered several possible fields of application, among them magnetocaloric cooling\cite{Annaorazov1996,Liu2012,Chirkova2016,Stern-Taulats2017}, heat assisted magnetic recording \cite{Thiele2003a,Bordel2012} antiferromagnetic memristor\cite{Marti2014},  spin valves \cite{Drozdz2018a} and recently also spin polarization detection\cite{Popescu2018}.

In our recent publication \cite{Witte2016} we reported a strain induced martensite-type transformation in highly strained epitaxial FeRh on W buffer layers. In contrast to the widely studied ordered phase, a chemically disordered alloy was investigated. The AF phase is highly sensitive to defects like anti-phase boundaries and disorder \cite{Uebayashi2006,Staunton2014} and thus FM magetic order prevails.  The disordered FM system reacts to the imposed strain during growth by transforming into a martensite structure with orthorhombic symmetry and a nanosized domain structure\cite{Witte2016}. Electronic structure  calculations suggest that the strain triggers a band-Jahn-Teller like lattice instability, responsible for the consequent transformation of the lattice. 

Here we present a systematic investigation of the strain adaptation mechanism in chemically disordered, near-equiatomic FeRh alloys as a function of epitaxial strain. For that, epitaxial thin film heterostructures are ideal model systems, as they allow for controlled growth of the interface with a well-defined orientation dependency between substrate and film. The subsequent investigation with X-ray reciprocal space mapping has identified an additional structural adaptation of the material, accessible only due to the well defined experimental model system.

\section{Experimental and computational details}
\subsection{Experimental procedures}

The $ ~$40\,-\,50\,nm thick W-V buffer layers were co-deposited by  dc-magnetron sputtering on epipolished MgO(001) single crystal substrates (SurfaceNET)  with a growth rate of $ ~$0.010\,-\,0.022\,nm/s at 350\,$^{\circ}$C  at an Ar pressure of 0.0011\,mbar. 
FeRh thin films of approximately 10\,nm thickness were deposited at ambient temperature  by using a Mini-electron-beam evaporator (Oxford Applied Research) allowing for the deposition of isotopically enriched  thin films for  use in  $^{57}$Fe conversion electron M\"ossbauer spectroscopy (CEMS).  The FeRh thin film growth was monitored \textit{in situ} with  RHEED (reflection high energy electron diffraction) and the observed  streak patterns confirmed  the epitaxial growth for all samples (see Supplementary Material\cite{SM}).
The chemical compositions of the  bilayers were determined with energy-dispersive x-ray spectroscopy (EDX): the obtained composition ($\pm 5$\,at\%) of the  W$_{1-z}$V$_{z}$\, layers and those of the Fe$_{1-x}$Rh$_{x}$\, layers are given in Tab.~\ref{tab:composition_lattice_constants_ferh_wv_rt}a) for five bilayers having buffer layers with different V content $z$. In the following, the samples are labeled with their $z$ parameter value.

Thickness $t$ of the individual layers of the heterostructure were determined from x-ray reflectometry (XRR) (see Tab.~\ref{tab:composition_lattice_constants_ferh_wv_rt}(b) ).
XRR and high resolution x-ray diffraction (HRXRD) measurements were performed with a Bruker D8 four-circle diffractometer, equipped with a Goebel mirror and a 4-bounce Ge(022) monochromator resulting in Cu-$K_{\alpha 1}$ radiation (0.154056\,nm). Reciprocal space maps (RSMs) were measured in co-planar geometry.
The films were isotopically enriched in $^{57}$Fe, enabling the investigation with element-specific $^{57}$Fe conversion electron M\"ossbauer spectroscopy (CEMS).  The High resolution transmission electron microscopy (HRTEM) study was carried out using a FEI Titan  80-300 electron microscope, operated at
an accelerating voltage of 300\,kV, featuring a CEOS image spherical aberration corrector. The  cross-section specimens were prepared using a focused ion beam (FIB) FEI Strata 400 S instrument. The first cutting step was performed with 30\,kV Ga$^+$ ions and the final polishing steps were done at 5\,kV and 2\,kV.

 \begin{table}[htb]\centering% add [H] placement to break table across pages
\caption{(a) Results of chemical analysis  of the W$_{1-z}$V$_{z}$/Fe$_{x}$Rh$_{1-x}$ bilayers by energy-dispersive X-ray spectroscopy (EDX). (b) Results of x-ray reflectometry (XRR) and high resolution x-ray diffraction (HRXRD), presenting the thickness $t$ of both layers, the out-of-plane lattice constant of the FeRh layer $a^{\prime}$ and the in-plane lattice constant $a_{\mathrm{WV}}$ of the buffer layer.} 

\begin{ruledtabular}
 \begin{tabular}{cc|cccc}

  \multicolumn{2}{c|}{(a) EDX} & \multicolumn{2}{c}{(b) XRR} & \multicolumn{2}{c}{ HRXRD}\\
 		\hline	

     $z$   &      $x$        &  $t_{\mathrm{FeRh}}$   (nm) & $t_{\mathrm{WV}}$ (nm)   & $a^{\prime}$ (\AA) &   $a_{\mathrm{WV}}$ (\AA) \\
     \hline
     	
      1.00 &   0.57(5)	   &12.5 & 54  &     2.672(2)  &  3.014(5)       \\   
   0.70(5) &  0.58(5)	  & 11.3 & 42   &      2.672(2)    & 3.030(5)  \\ 
   0.57(5) &   0.44(5)	  &  8.6 & 43 &       2.680(2)  &  3.066(5)  \\ 
   0.32(5) &    0.44(5)	  &  8.0 & 51 &       2.680(2)    &  3.108(5)    \\ 
   0.00 &   0.52(3)	      &  13.0& 50 &  2.655(2) &  3.151(5) \\      
   
  \end{tabular}
\end{ruledtabular}
\label{tab:composition_lattice_constants_ferh_wv_rt}
 
 \end{table}

\subsection{Computational details}

To obtain the binding surfaces for the ferromagnetic (FM) and paramagnetic (PM) phases, we carried out total energy calculations in the framework of density functional theory (DFT) using the Vienna Ab-initio Simulation Package (VASP).\cite{VASP1} Here, wavefunctions of the valence electrons are expressed in terms of a plane wave basis set, taking advantage of the projector augmented wave (PAW) approach,\cite{VASP2} to account for the interaction with the core electrons.
We took the PAW-potentials from the standard database generated for PBE-GGA (Perdew-Burke-Ernzerhof-General Gradient Approximation)\cite{cn:Perdew96}
which consider explicitly the electronic configuration
$2p^63d^74s^1$ for Fe  and $4p^64d^85s^1$ for Rh (versions of Sep.\ 2000).
To guarantee accurately optimized structures, the cutoff energy was chosen to be 366.5\,eV.

In order to keep the computational effort tractable, we made a few simplifying assumptions: Our calculations assume collinear magnetic moments for the FM and the PM phase, neglecting spin-orbit interactions. The PM phase is represented by a static arrangement of collinear Fe-spins with equal distribution of both directions. Statistic disorder is approximated in the framework of special quasi-random structures (SQS) \cite{Zunger1990}, which optimizes the elemental pair correlation functions of a disordered system in a small simulation cell.

To represent chemically and magnetically disordered structures on the same footing,
we employed as in our previous work\cite{Witte2016}
a cubic SQS of the type $A_2BC$ with 32 atoms which was previously published
by Jiang\cite{cn:Jiang09Acta}.  Averaging over the atomic occupation of the sites, this super cell
yields a bcc coordination for each atom. As epitaxial strain reduces the cubic symmetry, we must furthermore apply the distortion to all nonequivalent directions and average over these configurations.

In our calculations, we placed Rh on the $A$-sites of this SQS, as it is 
spin-polarized in the FM structure but carries negligible magnetic moments in the PM structure.
The $B$ and $C$ sites were occupied with Fe ions, which were spin-polarized in the same direction in the FM case. To model the PM structure, we occupied the $B$- and $C$-sites with Fe ions with positive and negative collinear magnetic moments, respectively.

Since the distribution of the atoms in the SQS super cell does not obey cubic symmetry, imposing a  tetragonal distortion
with the compressed $c$-axis oriented parallel  to each of the three Cartesian axes leads to three 
inequivalent configurations. To obtain a representative binding curve $E(c/a)$, the energies were thus averaged over all configurations
obtained in this way. 

We started from the cubic case and increased the distortion stepwise towards the maximum distortion, starting for each step from the optimized structures of the previous run.
After this, commencing now with the largest tetragonal distortion, we performed a second set of calculations  approaching again the cubic state by stepwise decreasing $c/a$. This quantifies the
non-reversibility arising from the ionic relaxation process, which tends to get stuck in local minima on the
binding surface. This procedure (see Supplementary Material for a graphical illustration\cite{SM}) was performed for both magnetic states at a given $c/a$.

For each fixed value of $c/a$, we carried out alternatingly full optimizations of atomic positions, interatomic forces and the volume of the simulation cell. We employed a convergence threshold of 0.01 eV/\AA for the forces and $10^{-7}\,$eV for the total energy, respectively. Electronic convergence was assumed when the energy between two consecutive electronic steps fell below $10^{-7}\,$eV.
For optimization, a $k$-mesh of 4$\times$4$\times$4 points was used together with the Brillouin zone
integration method of Methfessel and Paxton\cite{cn:Methfessel89} with a smearing of $\sigma=0.1\,$eV.
Finally, total energies were obtained with
a $k$-mesh of 6$\times$6$\times$6 points using the tetrahedron method with Bl\"ochl corrections for Brillouin zone
integration.\cite{cn:Bloechl94}

 \section{Results} 
 
\subsection{X-ray diffraction and reflectometry}

\begin{figure}[htb]\centering
\includegraphics[width=0.9\columnwidth]{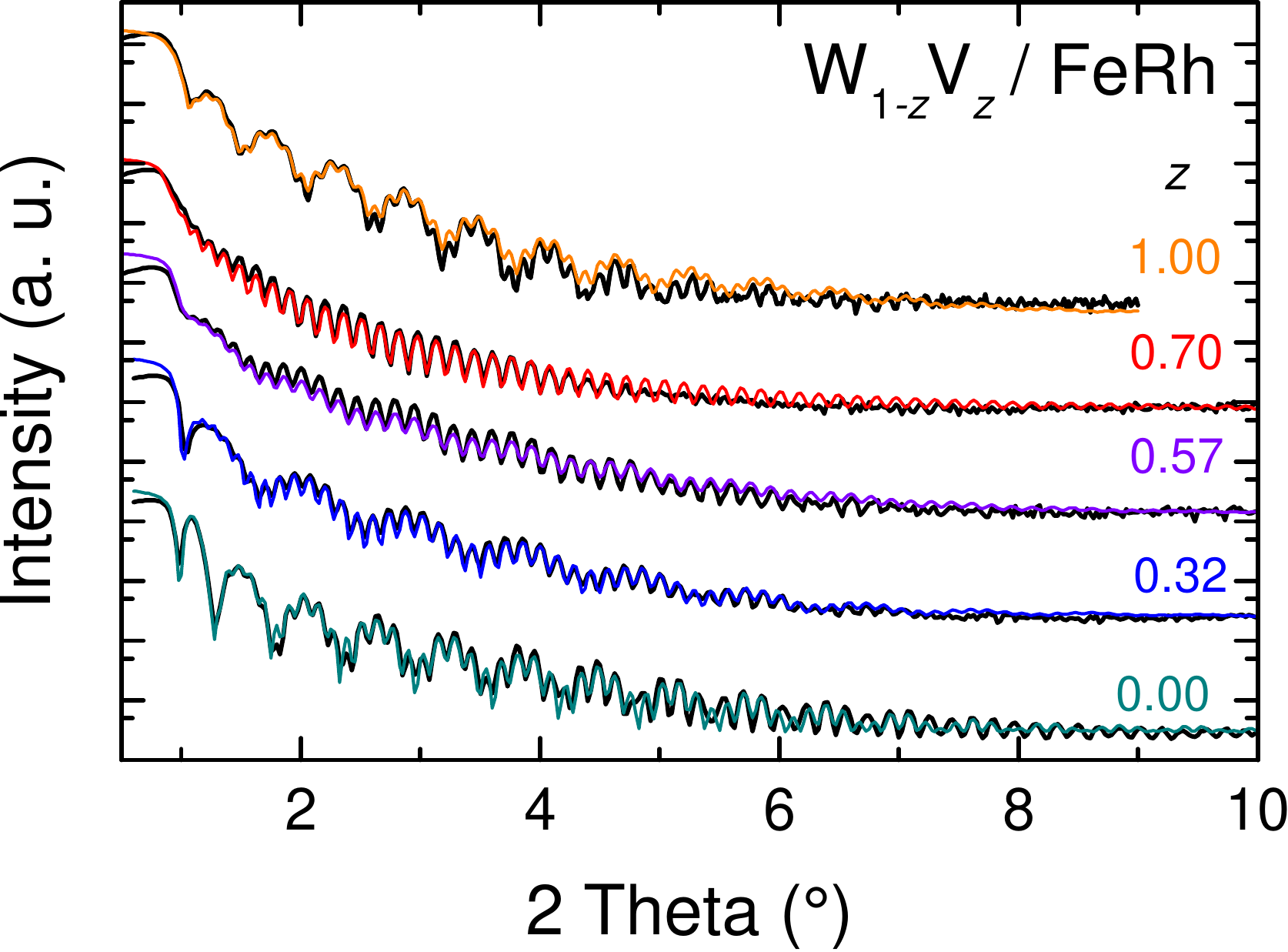}
\caption{X-ray reflectometry curves of FeRh/W$_{1-z}$V$_{z}$ bilayers. The obtained thicknesses of the respective layers are given in Tab.~\ref{tab:composition_lattice_constants_ferh_wv_rt}(b).}
\label{fig:xrr_ferh_wv_rt}
\end{figure}

XRR patterns of the five samples are presented  in Fig.~\ref{fig:xrr_ferh_wv_rt}. The observed Kiessig oscillations with dissimilar periodicity can be attributed to the two different layers of  the heterostructure. A clear distinction between the  oscillation stemming from the individual metal layers becomes difficult for the samples with  intermediate W$_{1-z}$V$_{z}$ composition. This is related to the difference in (electron) density between the two layers which provides the scattering contrast, and which is vanishing for the sample with $z=0.7$. 

\begin{figure}[htb]\centering
\includegraphics[width=0.9\columnwidth]{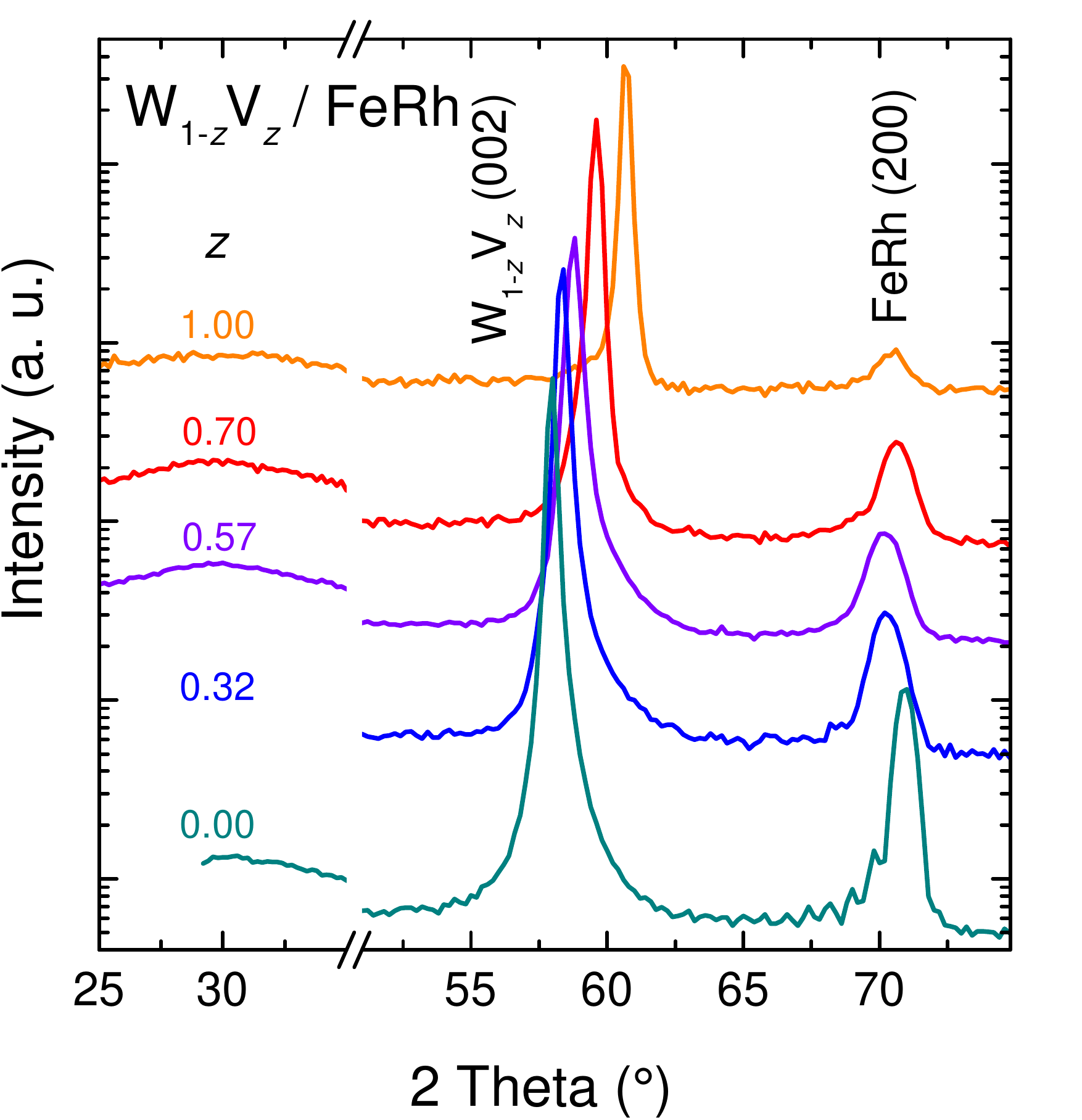}
\caption{Results of high resolution X-ray diffraction. (a) HRXRD patterns of FeRh/W$_{1-z}$V$_{z}$ bilayers grown at RT. The determined FeRh and W$_{1-z}$V$_{z}$ out-of-plane lattice parameters $a^{\prime}$ and $a_{\mathrm{WV}}$, respectively, are given in Tab.~\ref{fig:xrd_ferh_wv_rt}.}
\label{fig:xrd_ferh_wv_rt}
\end{figure} 

\begin{figure}[htb]\centering
    ~~~~\includegraphics[width=1\columnwidth]{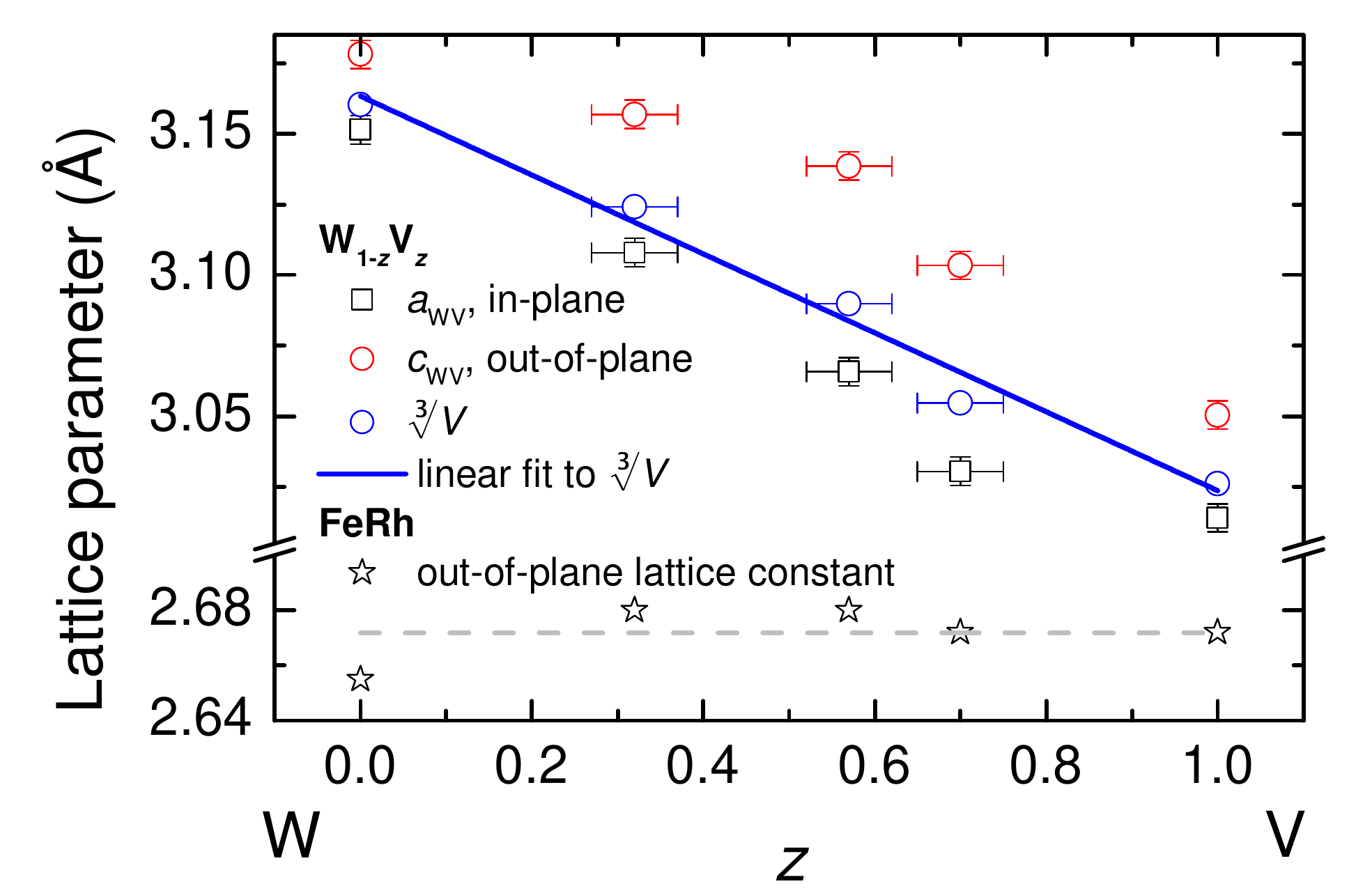}~~~~
 \caption{Synopsis of lattice parameters of W-V buffer layers and FeRh film in out-of-plane direction. 
  In- and out-of-plane lattice parameters $a_{\mathrm{WV}}$ and $c_{\mathrm{WV}}$, respectively, as well as the volume averaged lattice parameter $\sqrt[3]{V}$ of W$_{1-z}$V$_{z}$ buffer layers as a function of the V-content $z$. The solid line is a linear fit to $\sqrt[3]{V}$. In the lower part of the figure the FeRh out-of-plane lattice constant is plotted. The dashed line is a guide to the eye.}
 \label{fig:wv_structure}
\end{figure}

The results of the structural investigation with HRXRD are discussed first for the W$_{1-z}$V$_{z}$ buffer layers as they provide the basis for all further analysis. Fig.~\ref{fig:xrd_ferh_wv_rt} displays standard $\theta/2\theta$ measurements and the obtained out-of-plane lattice parameters $c_{\mathrm{WV}}$ are plotted as function of $z$ in Fig.~\ref{fig:wv_structure}, while the likewise presented in-plane lattice parameters $a_{\mathrm{WV}}$ are determined from measuring RSMs of asymmetric W$_{1-z}$V$_{z}$ (013) reflections (not shown).

Both parameters, $a_{\mathrm{WV}}$ and $c_{\mathrm{WV}}$, decrease systematically with increasing V content $z$. It is discernible  that in the medium composition region, the two values deviate stronger from each other, which is equivalent to an increasing tetragonal distortion of the bcc structure of the W-V alloys with $z$.
This implies a higher tolerance of the V-rich alloys compared to pure W with respect to epitaxial strain. This behavior is expected given that the epitaxial misfit to the underlying MgO (lattice spacing of 2.987\,\AA )  decreases with higher $z$ allowing for fully strained growth. At the same time the film material becomes less stiff (the elastic constants of V are about half of  those of W \cite{Featherston1963,Alers1960}) effectively reducing the elastic energy. Then, as $z$ increases towards pure V, the misfit is decreasing so that the  tetragonal distortion decreases again as well.  Additionally,  in Fig.~\ref{fig:wv_structure}  the volume averaged lattice parameter $\sqrt[3]{V}=\sqrt[3]{a_{\mathrm{WV}}^2c_{\mathrm{WV}}}$ is plotted, which averages out the effect of different degrees of strained growth assuming a constant unit cell volume upon tetragonal distortion. The observed linear behavior proves that the alloy thin films follow Vegard's rule, which is well known for  bulk W-V alloys \cite{Predel1998}. Here it is shown to be valid for epitaxial thin films and, moreover, that it can successfully be applied in tailoring the strain state of subsequently grown layers.

  The HRXRD measurements lead to another interesting observation: the position of the reflection attributed to the FeRh layer does not change with the composition  of the buffer layer, as can be seen from the diffraction patterns.
 This is a rather unexpected behavior of the FeRh layers, as in-plane and out-of-plane lattice constants  are usually linked; hence an influence would have been expected in some form. The lattice parameter values, corresponding to the $2\theta$ position, range around the values obtained for the FeRh films on W in the $Cmcm$ martensite phase (2.66\,-\,2.68\,\AA\, \cite{Witte2016}) rather than evolving towards the value expected for a bcc phase ($\approx 3$\,\AA\, \cite{Swartzendruber1983}), which is illustrated in the lower part  of Fig.\,\ref{fig:wv_structure}. Furthermore, our diffraction data prove by the absence of a
superstructure reflection in the lower angle regime
 that the room temperature deposition successfully suppressed the chemical ordering.

\begin{figure}[htb]\centering
\includegraphics[width=\columnwidth]{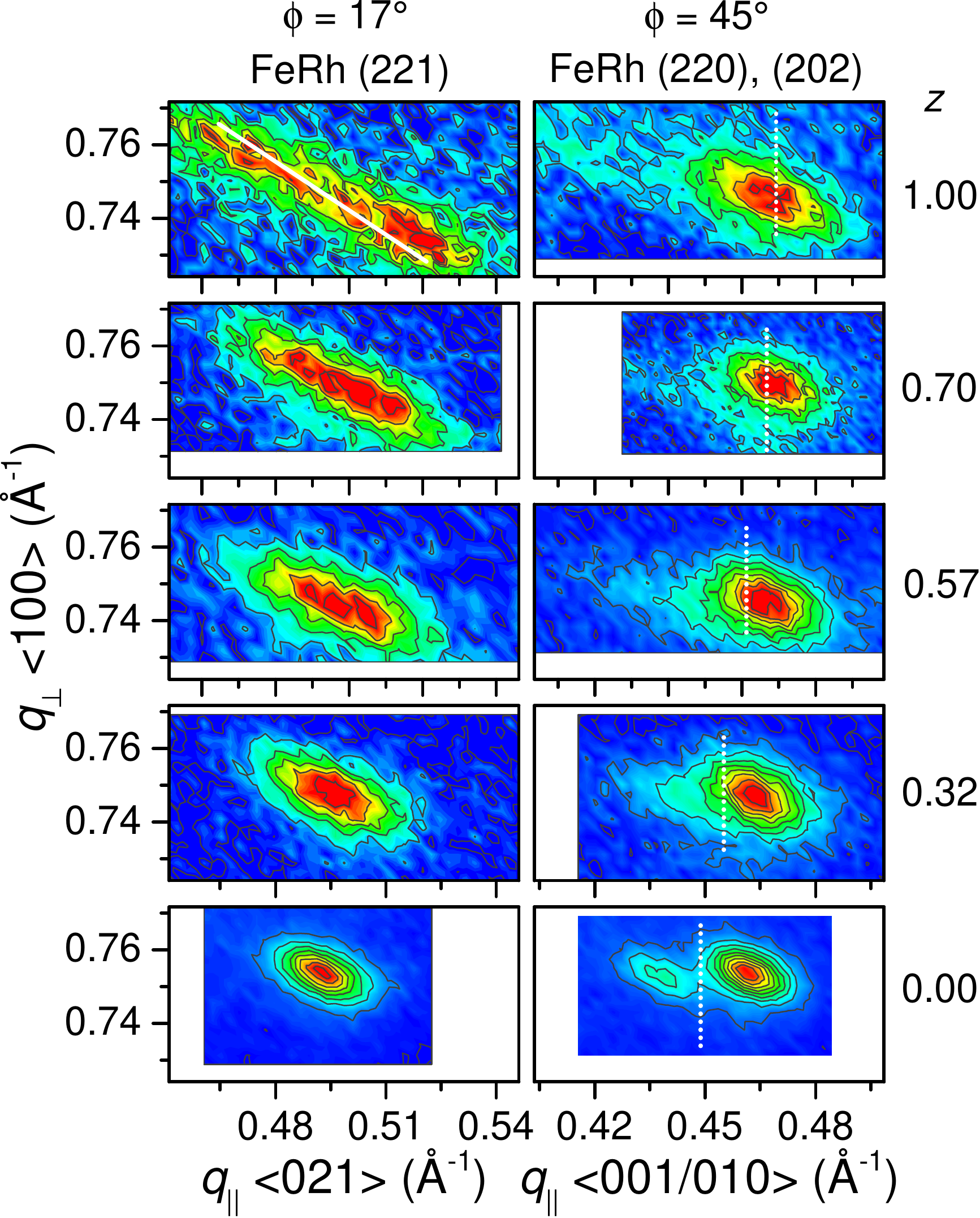}
\caption{Reciprocal space maps of FeRh/W$_{1-z}$V$_{z}$ bilayers. The left panels present measurements of the (221)$_{Cmcm}$ reflection: with  increasing V content $z$ the reflection becomes smeared out. The white bar in the topmost map is perpendicular to the scattering vector $q$, see text for detail. The right panels show the evolution of the (220),\,(202) reflection pair. The dotted white bar indicates the position of the in-plane scattering vector corresponding to the lattice of the W-V buffer layer.}
\label{fig:rsm_ferh_wv_rt}
\end{figure}

  RSMs of asymmetric reflections shown in Fig.~\ref{fig:rsm_ferh_wv_rt} can be exploited to shed light on the structure of the FeRh layers. Presupposing that the  symmetric $\theta/2\theta$ measurements yield lattice constants  consistent with the $Cmcm$ martensite phase in all the samples, asymmetric reflections or regions of reciprocal space were investigated  which permit  a clear distinction between a  tetragonal or the lower symmetry orthorhombic $Cmcm$ lattice, as seen in the figure.

  The RSMs of the (221) reflection (in $Cmcm$ notation) %, which is absent for a bcc or bct structure, 
  are shown in the left panels, while the (220),\,(202) pair is given in the right panels.  The  (220),\,(202) reflections would have counterparts in a possible cubic or tetragonal structure, but the splitting  is a direct sign of the two different in-plane lattice constants typical for an orthorhombic structure, as discussed in Ref.\,\onlinecite{Witte2016}.
 The observed reflections and lattice constants thus suggest the orthorhombic $Cmcm$ structure. The center position of the (221) reflections in reciprocal space does not decisively change with W$_{1-z}$V$_{z}$ composition. In contrast,  the width of the reflection changes dramatically. With decreasing in-plane lattice constant of the buffer layer, it becomes increasingly smeared out. This broadening is oriented approximately perpendicular to the scattering vector $q$, as indicated by the white bar $\perp\,q$ in the topmost panel. Such a broadening in this direction of reciprocal space  corresponds to a mosaic spread of the crystal planes, i.e. a tilting of the unit cells \cite{Spiess2009}. This information is intrinsically directional: that means it actually shows a mosaic spread, or tilting of the planes in the direction of the in-plane components of the chosen reflection $<h21>$,  selected by the azimuthal orientation $\phi$ of the sample. 

In order to be able to distinguish whether the observed mosaicity is isotropic or itself directional (meaning that the crystals or domains are tilted in certain specific crystallographic directions), it is necessary  to investigate reflections with different in-plane orientations. Indeed the measurement of the (220),\,(202) pair of reflections presented in the right panel of Fig.~\ref{fig:rsm_ferh_wv_rt}, provides evidence for directional (or anisotropic)  tilting. It is obvious that with decreasing lattice constant of the buffer layer (indicated by the white dotted bar), starting from pure W, the (220) reflection (left maximum) progressively smears out to such an extent that only a diffuse intensity distribution remains. In contrast, the (202) reflection is well defined, with only little broadening. These observations can be interpreted as follows:

 \begin{figure}[t]\centering
\includegraphics[width=1\linewidth]{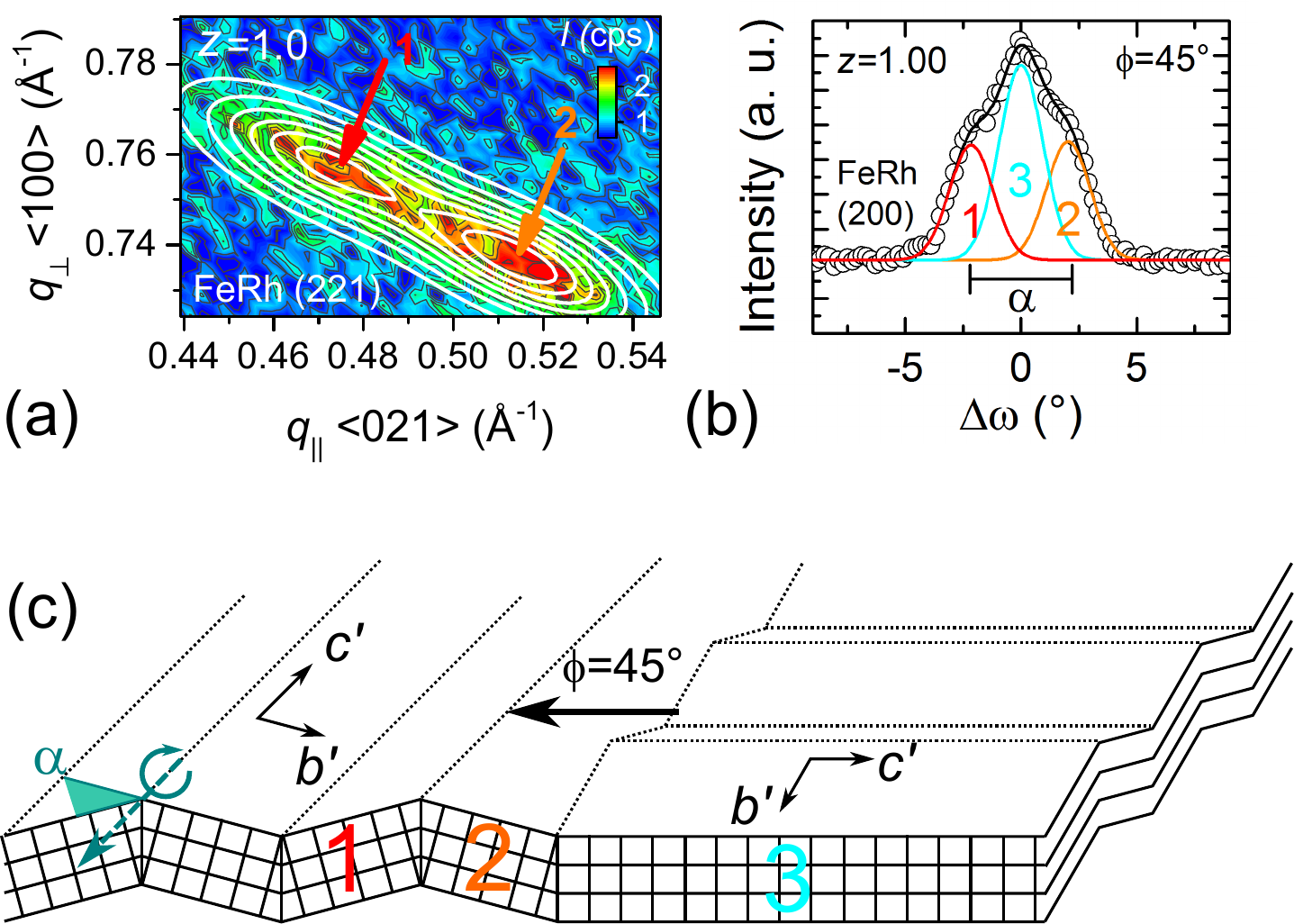}
\caption{(a) RSM of the FeRh (221) reflection for the film grown on pure V. The broad reflection can be reproduced with  two Gaussian peaks, indicating that the structural domains are tilted with a well defined tilt angle against each other. (b) Rocking curve of the FeRh (200) reflection, measured with an azimuthal angle $\phi$ of 45$^{\circ}$. The data can be fitted with three  individual Gaussian peaks (identical width). The labeling relates them to the differently oriented domains shown in (c), which presents a sketch of the proposed adaptive nanostructure, including, on top of the 90$^{\circ}$ in-plane rotated variants, also the additionally tilted domains along the $b^{\prime}$ direction. They can be attributed to the three components observed in the rocking curve. The tilt angle  $\alpha$ is also drawn, while the tilting is exaggeratedly displayed, the actual value of  $\alpha$\, is $\approx 4.2^{\circ}$.  The  interfaces or habit planes between the domains are drawn with dotted lines. This is meant to indicate that the real habit planes cannot be identified, only the orientation of the crystalline domains is known. See text for more details.}
\label{fig:rsm_221_wv_rt}
\end{figure} 

 The broadening of the (220) reflection indicates that in this particular crystallographic (in-plane) direction, namely along the $b^{\prime}$ axis, the tilting of the crystallographic domains is strongest,  while the lattice along the $c^{\prime}$ axis is undisturbed. The (221) reflection, having an in-plane component in both directions, thus also shows  signs of tilting but not to the same extent as the (220) reflection.
The (221) intensity plot  can be even separated  into two overlapping maxima, which can be fitted with two 2D Gaussian peaks as  shown in Fig.~\ref{fig:rsm_221_wv_rt}(a). This is an important observation as it implies a well defined tilt angle between the two crystallographic domains and not just  a broad mosaic spread.

This structural feature should  also be observed in rocking curves of symmetric reflections as a function of the in-plane rotational orientation $\phi$.  Fig.~\ref{fig:rsm_221_wv_rt}(b) displays the rocking curves of the  reflection coming from the FeRh (200) planes  for an azimuthal angle $\phi$ of 45$^{\circ}$ ($\phi$ of 0$^{\circ}$ corresponds to the direction of the principal axes of the bcc W-V buffer layer), hence parallel to the $\langle 010 \rangle$,\,$\langle 001 \rangle$  directions (see orientation relations in the sketch in  Fig.~\ref{fig:rsm_221_wv_rt}(c)). Indeed the $\phi=45^{\circ}$ measurement shows a strong broadening and the curve can be deconvoluted into three peaks, the two outer ones corresponding to the  domains tilted in this azimuthal direction (labeled 1 and 2). These maxima are thus shifted away from the center position, while the center peak corresponds to the domains oriented in the perpendicular in-plane direction (labeled 3). From the position of the outer peaks a maximum tilt angle of  $\alpha\approx4.2^{\circ}$  between the two domains can be determined.

 Moreover, the observation that the structural domains are tilted in $b^{\prime}$ direction, which can also be seen as an in-plane rotation of the domain orientations around the $c^{\prime}$ axis, is very intriguing. %, given that the $c^{\prime}$ axis has been identified as the pseudo-{hcp} axis of the $Cmcm$ structure. 
 Whereas the $c^{\prime}$ axis itself is, as evidenced by the RSM, not affected from the tilting and remains parallel to the film plane.

 A tentative sketch of the  nanostructure, illustrating the tilted structural domains, is shown in Fig.~\ref{fig:rsm_221_wv_rt}(c). In the graph  each peak  in the RSM is referred to one  (arbitrarily chosen) orientation of the tilted domains. The figure also illustrates the tilt angle $\alpha$ between the domains, which corresponds to a rotation of the lattice planes around the $c^{\prime}$-axis.
 The habit planes between the tilted and in-plane rotated domains have not been identified, therefore they are only indicated by the dotted lines.

\subsection{Transmission electron microscopy} 
 
  \begin{figure*}[hbt]\centering
\includegraphics[width=1\textwidth]{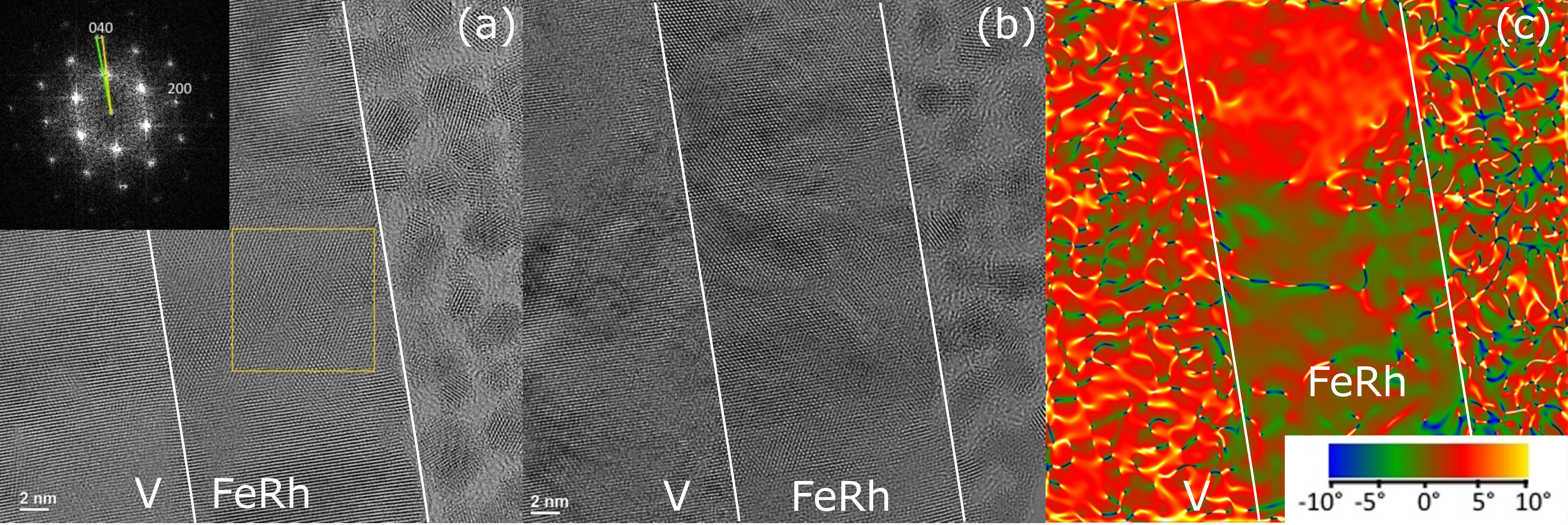}
\caption{HRTEM.(a) Cross sectional view through the heterostructure, insets shows a Fast-Fourier-Transform (FFT) from the framed region. The FFT shows two patterns with slight in-plane rotation, as seen by the splitting of the 040 reflection. The two diffracting grains are placed along the viewing direction, and the Moir\'e fringes in the high resolution image evidence the misorientation. (b) HRTEM of a different area of the cross section, here the  misorientation can be observed for two neighboring grains, as seen in the Geometrical phase analysis (GPA) in (c) of the same area as in (b). Here the colour code refers to a rotation of the lattice planes.}
\label{fig:tem}
\end{figure*}

 High resolution transmission electron microscopy (HRTEM) allows to study the in-plane rotation of the $c$-axis oriented domains on a local nano-scale. Fig.\,\ref{fig:tem} (a) shows such a cross sectional HRTEM micrograph. From the left to the right, one can identify the V buffer layer and  then the epitaxially grown FeRh buffer layer. Fast Fourier transform (FFT) was performed from the framed area in the FeRh layer, where the lattice fringes are modulated by the varying brightness (Moir\'e effect). From the FFT, it can be clearly seen that the 020 reflections split with a small misorientation angle of about 3.6$^{\circ}$ (rotation around the $c$-axis, which is the zone axis in this image). This means that along the viewing direction, there are overlapped parts or grains with a small rotation angle between each other. 
 
 In different areas of the prepared cross section lamella, such misorientation can be observed for directly neighboring grains, as shown in Fig.\,\ref{fig:tem}(b-c). By using a geometrical phase analysis (GPA) \cite{Hytch1998}, the rotation of the lattices can be quantified for the different areas in the HRTEM image. The upper part coded by red color is misoriented by about 3.8$^{\circ}$ compared to the lower part in the epitaxial layer, which is in agreement with the analysis of FFT for the previous HRTEM image.

This locally observed  misorientation  corresponds directly  to the grains  of type 1 and 2 (referring to Fig.\,\ref{fig:rsm_221_wv_rt}), which were seen on a macroscopic scale by x-ray diffraction. The determined misorientation angles of about 3.6 to 3.8$^{\circ}$ agree well with the value determined from the rocking curves.

 \subsection{Conversion electron M\"ossbauer spectroscopy}
The conclusions from  x-ray diffraction that the actual crystallographic structure is  unchanged throughout the sample series is supported in addition by CEMS, which provides a local spectroscopic view on the environment around the $^{57}$Fe probe nuclei.
The obtained spectra (see Supplementary Material\cite{SM}) are basically unaltered as function of the buffer layer lattice constant. All spectra can  be represented with a narrow doublet with a quadrupole splitting of $\Delta E_Q $ $\sim$0.13\,-\,0.16\,mm/s and an isomer shift  $\delta$ of $\sim$0.06\,-\,0.7\,mm/s (see Tab.\,\ref{tab:cems_thickness_ferh_wv_rt} with fitting results). All parameters are basically identical within the error margin, pointing towards an effectively similar local environement. 

This finding clearly is different to our previous work \cite{Witte2017},  where we investigated differently strained FeRh thin films with partial chemical ordering. In that case, two subspectra were observed, one doublet attributed to the $Cmcm$ phase and one sextet stemming from a distorted bct phase with partial chemical ordering. 
  Hence, the present CEMS results yield  valuable local spectroscopic evidence which supports the hypothesis that the local crystallographic structure of the FeRh films remains  close to the  $Cmcm$ martensite structure\cite{Witte2016}, showing no sign of secondary phases, in agreement with the analysis of the HRXRD measurements.

 \begin{table}[htb]\centering% add [H] placement to break table across pages
\caption{M\"ossbauer hyperfine para\-meters obtained by fitting quadrupole doublets to the room temperature CEMS spectra: isomer shift $\delta$, quadrupole splitting $\Delta E_Q $, Lorentzian linewidth $\Gamma $, Gaussian broadening $\sigma $ (all given in mm/s).   $\delta$ is given with respect to bcc-Fe at room temperature.} \label{tab:cems_thickness_ferh_wv_rt}
\begin{ruledtabular}
 \begin{tabular}{ccccc}

        $z$ & $\delta$ & $\Delta E_Q $ & $\Gamma $ &$\sigma $ \\
      
    \hline  
        1  & 0.068(5) & 0.160(4) & 0.275(6) & 0.26(1)\\
 	   0.7 & 0.072(5) & 0.132(3) & 0.24(1) & 0.26(4)\\
       0.57& 0.072(5) & 0.145(2) & 0.24(3) & 0.26(5)\\
      0.32 & 0.071(5) & 0.141(2) & 0.23(2) & 0.25(2)\\
        0  & 0.063(5) & 0.148(5) & 0.280(5) & 0.25(1) \\
        
   \end{tabular}
\end{ruledtabular}
 \end{table}

\subsection{Density functional theory calculations}

 \begin{figure}[hbt]\centering
\includegraphics[width=1\linewidth]{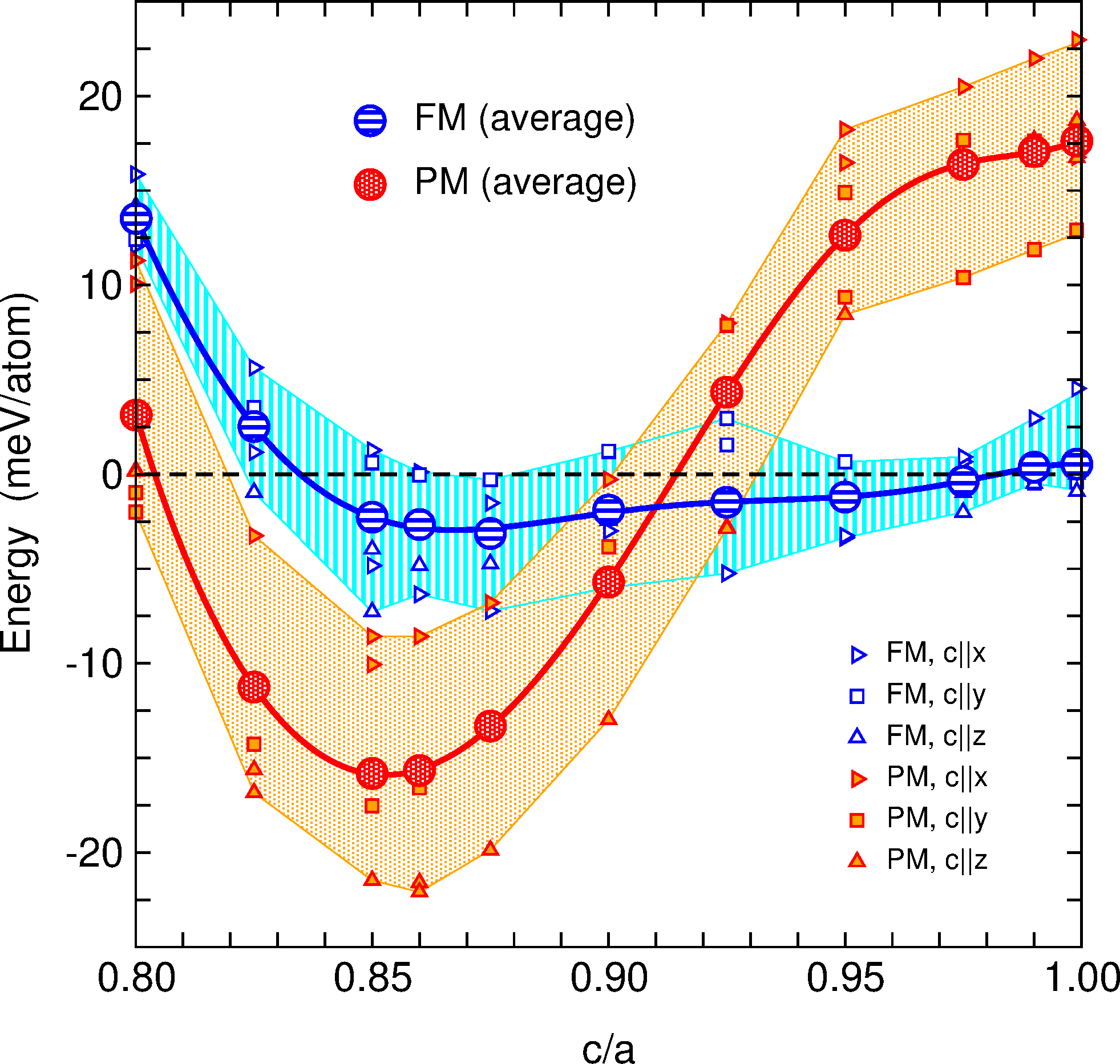}
\caption{Energy as function of the tetragonal distortion $c/a$ for chemically disordered FeRh
from density functional theory. The blue and red circles denote the average energy of all calculations
in the FM and PM state carried out for a given $c/a$, while the smaller symbols represent the
individual values obtained from a consecutive deformation
of the structure along the three Cartesian axes.
The shaded regions illustrate the range between the minimum and maximum energy configurations
for each magnetic state quantifying the uncertainty range for the calculated values.
The lines are only guides to the eye.}
\label{fig:dft}
\end{figure} 

To complement the microscopic understanding of the diffraction data, we carried out first-principles calculations in the framework of density functional theory calculations, obtaining the total energy landscape for tetragonally distorted and chemically disordered FeRh with ferromagnetic order (FM) and in a configuration with static magnetic disorder (referred to as paramagnetic, PM), see Fig.\,\ref{fig:dft}.
The volume of the cell and atomic positions have been fully optimized, while the tetragonal distortion was fixed in the range $0.8\leq c/a\leq 1$ which corresponds to the epitaxial boundary conditions in the experiment.
For the cubic  A2 phase at $c/a=1$  we find that FM order is with $17\,$meV/atom substantially lower in energy than
the corresponding PM A2 configuration% in agreement with the experimental observation of ferromagnetic order in this case
. Afterwards we increased the tetragonal distortion step-wise, starting from the positions and magnetic configuration of the previous run, subsequently optimizing atomic positions and cell volume once again.
Tetragonal deformation affects the energy of the FM structure only
slightly. The energy landscape remains remarkably flat, forming a local minimum around $c/a=0.87$. Its energy is,
however, only $3.5\,$meV below the A2 FM, which is at the edge of the accuracy of our calculations.
A similarly flat energy landscape has been revealed for the magnetic shape memory system
Fe$_{70}$Pd$_{30}$.\cite{cn:Opahle09,cn:Weiss11PRL,cn:Gruner11PRB}
This latter system has a comparable valence electron ratio $e/a=8.6$ ($e/a=8.5$ for Fe$_{50}$Rh$_{50}$)
and undergoes a martensitic transformation in the respective composition range.
Larger deformations, finally, increase the energy significantly in the FM state of Fe$_{50}$Rh$_{50}$.

The picture looks entirely different for the PM case. Starting at $c/a=1$ with a slight decrease as in the FM case,
the slope increases significantly for distortions larger than 5\,\%, reaching its minimum
around $c/a=0.85$ being now 13\,meV/atom below the tetragonally distorted FM and 16.5\,meV/atom below
the cubic A2 structure. We find a crossover between the two magnetic structures at around $c/a=0.9$.
This is indicative of the transformation from a FM cubic or tetragonal phase to
the paramagnetic phase with orthorhombic $Cmcm$ symmetry, which we identified earlier as the new
potential ground state of epitaxially
strained disordered FeRh\cite{Witte2016}. In addition
the energies agree well with our previous results obtained for isolated calculations at
$c/a=1$ and $c/a=0.87$ for both the FM and PM cases. 
In addition, we have also reversed the deformation-relaxation-procedure, in terms of a step-wise decrease of $c/a$,
starting from the largest tetragonal distortion. The final configurations obtained for $c/a=1$ differ from the
initial configurations only moderately by a few meV/atom, the initial order of the phases is restored. This proves that the deformation-induced magnetostructural transition, which we observe at $c/a=0.9$, is essentially reversible
apart from positional disorder which traps the system in local minima on the binding surface and
can thus lead to enhanced hysteresis.

\section{Discussion}

Directional tilting of epitaxial layers has been already described in the literature as a strain or epitaxial misfit relaxation process \cite{Ayers1991}. It has been observed in the heteroepitaxy of  e.g. ferroelectric materials \cite{Kim2006,Lee2000b} but as well in compound semiconductor thin films \cite{Schowalter1990}. However, in the case of the present strained FeRh thin films, it is not immediately apparent why such a mechanism should appear.

 Our discussion begins with the FeRh thin film on the elemental W buffer layer: for that situation, the disordered bcc A2 phase has a  lattice spacing closest to  the W buffer layers (+6\%), while  the  disordered A1 phase with face-centered-cubic (fcc)  symmetry has a huge epitaxial misfit of $\approx +20$\%. That is why the A2 phase is chosen as a reference   for the films grown on W as presented in  Ref.\,\onlinecite{Witte2016}, despite the fact that the A1 phase is known to be thermodynamically more stable than the A2 phase \cite{Ohnuma2009}. However,  the system avoids the  A2-like bct phase, by growing in the obviously more stable close packed $Cmcm$ structure.

\begin{figure}[htb]\centering
\includegraphics[width=0.8\linewidth]{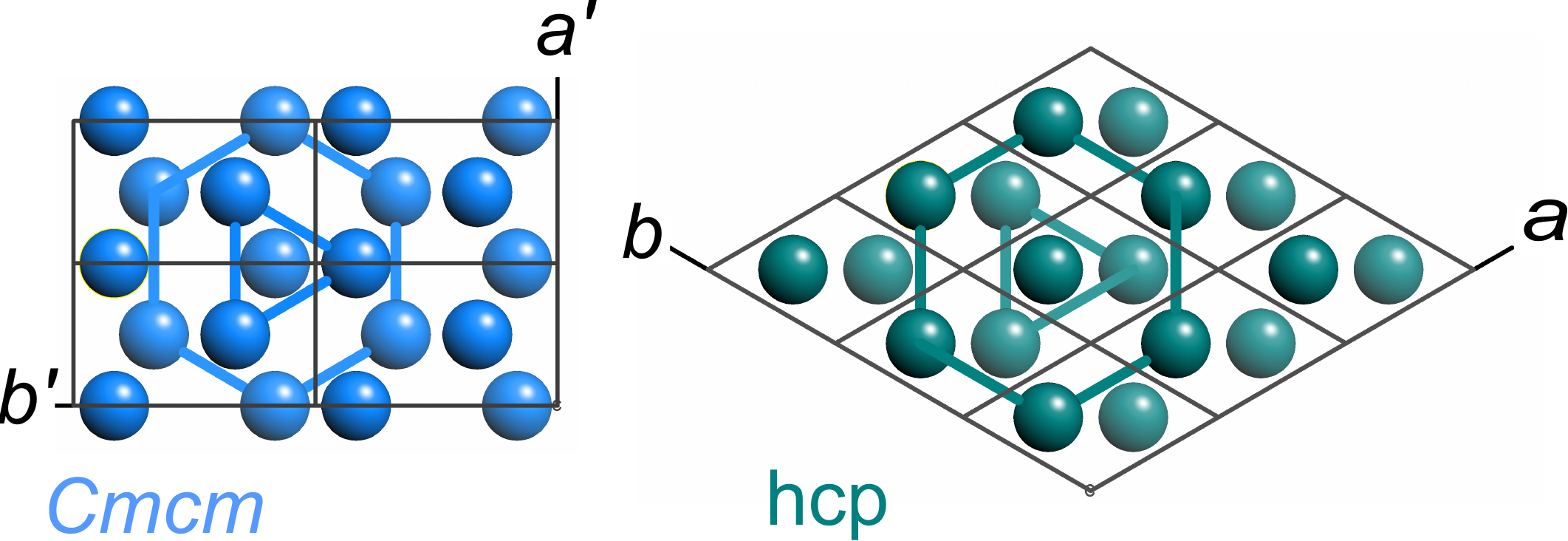}
\caption{Comparison of $Cmcm$ and hcp structure. Both structures are shown along their (pseudo) hexagonal $c$ axis.}
\label{fig:comp_hcp_cmcm}
\end{figure}

The structural arrangement of the $Cmcm$ lattice bears similarities to a hexagonal close packed (hcp) structure. This can be understood by looking along the $c^{\prime}$ axis as illustrated in Fig.\,\ref{fig:comp_hcp_cmcm}. The resulting pattern resembles a threefold axis with an ABA stacking, typical of a hcp structure. In this comparison, the hexagonal $c$-axis of a possible hcp structure would be oriented in the plane, while the <110>$_{hcp}$ direction represents the out-of-plane direction. 
Nevertheless, the complete set of crystallographic reflections unequivocally proves  an orthorhombic symmetry of the crystallographic structure \cite{Witte2016,Witte2016a} rather than a hexagonal one. It can be thus speculated that the pseudo-closed packed $Cmcm$ arrangement, including the nanosized domain pattern, is the result of a compromise between  a close-packed structure and  fulfilling the epitaxial boundary condition. In the calculation, this is expressed by the local minimum at $c/a\approx0.87$, where we previously identified the $Cmcm$ structure. To  that respect the W buffer layers  matches well to the average in-plane lattice constants of the equilibrium $Cmcm$ structure.

 In the case of the FeRh films on W-V buffers, the preset in-plane lattice parameter is reduced, so that the epitaxial misfit decreases with respect to the metastable A2 phase, reaching down to only +1\%. From this purely geometrical consideration, one thus expects a stabilization of the FM, cubic A2 phase on the V-rich and pure V buffer layers. 
 But as we can see from our calculations a PM $Cmcm$ phase with an effective $c/a\approx0.87$ as found in the experiments for the films on V, is significantly lower in energy than the corresponding cubic phase at $c/a=1$.
 
Moving now  from the  $Cmcm$ structure on pure W to the V-rich buffer layers, we find that the system tries to remain in the pseudo-close packed $Cmcm$ arrangement, instead of transforming towards a tetragonal distorted A2 phase, which is indeed energetically less favorable. But with decreasing lattice constant of the buffer layer, the $Cmcm$ structure does not perfectly  fit to the lattice constant of the buffer anymore. The mismatch is mostly due to the longer $b^{\prime}$ axis. The shorter $c^{\prime}$ lattice parameter now matches better to  the buffer layer and both eventually coincide. This is illustrated by the overlap of the white dotted bar (in-plane position of the buffer layer reflection) with the position of the FeRh (202) reflection in the case of the pure V sample (see Fig.\,\ref{fig:rsm_ferh_wv_rt}). However, the misfit in the $b^{\prime}$ direction drastically increases, which ultimately is  the reason for the observed directional tilting of the crystallographic domains along the direction of the $b^{\prime}$ axis. This additional adaptation mechanism allows the system to end up in the local minimum for the $Cmcm$ phase at $c/a\approx0.87$, although the epitaxial conditions should favor the cubic phase.

This interpretation suggests that the metastable close packed structures dominate in the entire range of in-plane  lattice constants investigated in these chemically disordered alloys and the expected FM A2 phase cannot be stabilized\footnote{At least in the investigated alloy concentration; alloys with higher Fe contents may be stabilized in the cubic structure.}. Further experiments exploring the strain adaptation behavior for smaller buffer layer lattice constants, towards an fcc structure, or even larger buffer layer lattice constants can be realized using Cr-V,\cite{Kaspar2014} Cu-Au\cite{Kauffmann-Weiss2014} or possibly Ta-W\cite{Franke} buffer layers, respectively.

The importance of chemical order (B2-order) for the strain adaptation behavior is illustrated by our recent results on FeRh thin films, which were grown at elevated temperature on W-V buffers \cite{Witte2017}. Here the thermal energy allows for  partial chemical ordering of the structure and in turn the strain adaptation mechanism is different. In that case a two-phase nanostructure consisting of the $Cmcm$ phase and a chemically ordered B2 phase is formed. This two-phase nanostructure adapts to the buffer layer lattice by varying the respective ratio of the two phases.

Together with the present results the findings on a whole display  a remarkable feature of the Fe-Rh materials system. Depending on the degree of chemical ordering the material is able to adapt to any preset lattice constant by either formation of a martensitic nanostructure, in effect a  fine tuning adjustment by the directional tilting of the martensitic domains or by the formation of a self assembled two phase nanostructure \cite{Witte2017} .

\section{Summary and Conclusion}
In summary, the results of this systematic investigation show  that the $Cmcm$ structure is stable over a wide  range of in-plane lattice parameters. This is  due to an adaptive mechanism based on the observed alternate tilting of the crystallographic domains in direction of the longer in-plane axis $b^{\prime}$, which leads to an  effectively reduced misfit in this particular direction. Adaptive structures do not only allow an accommodation of the strain, but also display a possibility for the system to avoid the thermodynamically unfavorable disordered A2 phase, despite the epitaxial condition, which clearly would favor the  latter phase. DFT calculations prove that the PM $Cmcm$ phase with a  $c/a\approx0.87$ is lower in energy than an FM cubic phase, and the observed tilting of the nano-domains allows the system to achieve this local energy minimum. Moreover, our calculations suggest the existence of a magnetostructural transition at $c/a\approx0.9$. 

As such this study provides new insights into the strain adaptation mechanism of chemically disordered Fe-Rh alloys and correlates the observed behavior to the free energy landscape of the system. This is discussed also with respect to the (partially) B2 ordered phases \cite{Witte2017}, which show an entirely different strain adaptation behavior, which includes the separation into a two phase nanostructure. The fact that a variety of strain adaptation mechanisms exist in one binary alloy, as function of degree of chemical order and strain,  is an interesting feature of the Fe-Rh system and may occur as well in other alloy systems in which chemical ordering tendencies are present.

With this study we have shown that epitaxial heterostructures consisting of alloy buffer layers and the functional material of interest can be employed to study strain adaptation mechanisms in a well defined  model system. This allows to vary strain and misfit in a large range and also gives access to directional strain adaptation mechanisms, which has proven  crucial for the present case study.

The importance of interfacial strain and strain adaptation mechanisms for modern functional materials such as  (magnetic) shape memory alloys is highlighted. However the formation of  heteroepitaxial interfaces also plays a role in energy storage materials \cite{Hao2012,Lee2015} and the proposed approach may be also successfully applied to these cases.

\section{Acknowledgement}
  The authors acknowledge funding by the Deut\-sche For\-schungs\-ge\-mein\-schaft via HA 1344/28-1 and GR 3498/3-2 (SPP1599). The experimental support by T. Scherer (KIT) is also acknowledged.

%\bibliography{library,theo}

\begin{thebibliography}{61}%
\makeatletter
\providecommand \@ifxundefined [1]{%
 \@ifx{#1\undefined}
}%
\providecommand \@ifnum [1]{%
 \ifnum #1\expandafter \@firstoftwo
 \else \expandafter \@secondoftwo
 \fi
}%
\providecommand \@ifx [1]{%
 \ifx #1\expandafter \@firstoftwo
 \else \expandafter \@secondoftwo
 \fi
}%
\providecommand \natexlab [1]{#1}%
\providecommand \enquote  [1]{``#1''}%
\providecommand \bibnamefont  [1]{#1}%
\providecommand \bibfnamefont [1]{#1}%
\providecommand \citenamefont [1]{#1}%
\providecommand \href@noop [0]{\@secondoftwo}%
\providecommand \href [0]{\begingroup \@sanitize@url \@href}%
\providecommand \@href[1]{\@@startlink{#1}\@@href}%
\providecommand \@@href[1]{\endgroup#1\@@endlink}%
\providecommand \@sanitize@url [0]{\catcode `\\12\catcode `\$12\catcode
  `\&12\catcode `\#12\catcode `\^12\catcode `\_12\catcode `\%12\relax}%
\providecommand \@@startlink[1]{}%
\providecommand \@@endlink[0]{}%
\providecommand \url  [0]{\begingroup\@sanitize@url \@url }%
\providecommand \@url [1]{\endgroup\@href {#1}{\urlprefix }}%
\providecommand \urlprefix  [0]{URL }%
\providecommand \Eprint [0]{\href }%
\providecommand \doibase [0]{http://dx.doi.org/}%
\providecommand \selectlanguage [0]{\@gobble}%
\providecommand \bibinfo  [0]{\@secondoftwo}%
\providecommand \bibfield  [0]{\@secondoftwo}%
\providecommand \translation [1]{[#1]}%
\providecommand \BibitemOpen [0]{}%
\providecommand \bibitemStop [0]{}%
\providecommand \bibitemNoStop [0]{.\EOS\space}%
\providecommand \EOS [0]{\spacefactor3000\relax}%
\providecommand \BibitemShut  [1]{\csname bibitem#1\endcsname}%
\let\auto@bib@innerbib\@empty
%</preamble>
\bibitem [{\citenamefont {Schlom}\ \emph {et~al.}(2007)\citenamefont {Schlom},
  \citenamefont {Chen}, \citenamefont {Eom}, \citenamefont {Rabe},
  \citenamefont {Streiffer},\ and\ \citenamefont {Triscone}}]{Schlom2007a}%
  \BibitemOpen
  \bibfield  {author} {\bibinfo {author} {\bibfnamefont {D.~G.}\ \bibnamefont
  {Schlom}}, \bibinfo {author} {\bibfnamefont {L.-Q.}\ \bibnamefont {Chen}},
  \bibinfo {author} {\bibfnamefont {C.-B.}\ \bibnamefont {Eom}}, \bibinfo
  {author} {\bibfnamefont {K.~M.}\ \bibnamefont {Rabe}}, \bibinfo {author}
  {\bibfnamefont {S.~K.}\ \bibnamefont {Streiffer}}, \ and\ \bibinfo {author}
  {\bibfnamefont {J.-M.}\ \bibnamefont {Triscone}},\ }\href {\doibase
  10.1146/annurev.matsci.37.061206.113016} {\bibfield  {journal} {\bibinfo
  {journal} {Annual Review of Materials Research}\ }\textbf {\bibinfo {volume}
  {37}},\ \bibinfo {pages} {589} (\bibinfo {year} {2007})}\BibitemShut
  {NoStop}%
\bibitem [{\citenamefont {del Alamo}(2011)}]{DelAlamo2011}%
  \BibitemOpen
  \bibfield  {author} {\bibinfo {author} {\bibfnamefont {J.~A.}\ \bibnamefont
  {del Alamo}},\ }\href {\doibase 10.1038/nature10677} {\bibfield  {journal}
  {\bibinfo  {journal} {Nature}\ }\textbf {\bibinfo {volume} {479}},\ \bibinfo
  {pages} {317} (\bibinfo {year} {2011})}\BibitemShut {NoStop}%
\bibitem [{\citenamefont {Khachaturyan}\ \emph {et~al.}(1991)\citenamefont
  {Khachaturyan}, \citenamefont {Shapiro},\ and\ \citenamefont
  {Semenovskaya}}]{Khachaturyan1991}%
  \BibitemOpen
  \bibfield  {author} {\bibinfo {author} {\bibfnamefont {A.~G.}\ \bibnamefont
  {Khachaturyan}}, \bibinfo {author} {\bibfnamefont {S.~M.}\ \bibnamefont
  {Shapiro}}, \ and\ \bibinfo {author} {\bibfnamefont {S.}~\bibnamefont
  {Semenovskaya}},\ }\href {\doibase 10.1103/PhysRevB.43.10832} {\bibfield
  {journal} {\bibinfo  {journal} {Physical Review B}\ }\textbf {\bibinfo
  {volume} {43}},\ \bibinfo {pages} {10832} (\bibinfo {year}
  {1991})}\BibitemShut {NoStop}%
\bibitem [{\citenamefont {{K. Bhattacharya}}(2004)}]{Bhattacharya2004}%
  \BibitemOpen
  \bibfield  {author} {\bibinfo {author} {\bibnamefont {{Kaushik
  Bhattacharya}}},\ }\href@noop {} {\emph {\bibinfo {title} {Microstructure of
  Martensite}}}\ (\bibinfo  {publisher}
  {Oxford University Press, Oxford},\ \bibinfo {year} {2004})\BibitemShut {NoStop}%
\bibitem [{\citenamefont {Kaufmann}\ \emph {et~al.}(2010)\citenamefont
  {Kaufmann}, \citenamefont {R{\"{o}}{\ss}ler}, \citenamefont {Heczko},
  \citenamefont {Wuttig}, \citenamefont {Buschbeck}, \citenamefont {Schultz},\
  and\ \citenamefont {F{\"{a}}hler}}]{Kaufmann2010}%
  \BibitemOpen
  \bibfield  {author} {\bibinfo {author} {\bibfnamefont {S.}~\bibnamefont
  {Kaufmann}}, \bibinfo {author} {\bibfnamefont {U.~K.}\ \bibnamefont
  {R{\"{o}}{\ss}ler}}, \bibinfo {author} {\bibfnamefont {O.}~\bibnamefont
  {Heczko}}, \bibinfo {author} {\bibfnamefont {M.}~\bibnamefont {Wuttig}},
  \bibinfo {author} {\bibfnamefont {J.}~\bibnamefont {Buschbeck}}, \bibinfo
  {author} {\bibfnamefont {L.}~\bibnamefont {Schultz}}, \ and\ \bibinfo
  {author} {\bibfnamefont {S.}~\bibnamefont {F{\"{a}}hler}},\ }\href {\doibase
  10.1103/PhysRevLett.104.145702} {\bibfield  {journal} {\bibinfo  {journal}
  {Physical Review Letters}\ }\textbf {\bibinfo {volume} {104}},\ \bibinfo
  {pages} {145702} (\bibinfo {year} {2010})}\BibitemShut {NoStop}%
\bibitem [{\citenamefont {Niemann}\ \emph {et~al.}(2012)\citenamefont
  {Niemann}, \citenamefont {R{\"{o}}{\ss}ler}, \citenamefont {Gruner},
  \citenamefont {Heczko}, \citenamefont {Schultz},\ and\ \citenamefont
  {F{\"{a}}hler}}]{Niemann2012}%
  \BibitemOpen
  \bibfield  {author} {\bibinfo {author} {\bibfnamefont {R.}~\bibnamefont
  {Niemann}}, \bibinfo {author} {\bibfnamefont {U.~K.}\ \bibnamefont
  {R{\"{o}}{\ss}ler}}, \bibinfo {author} {\bibfnamefont {M.~E.}\ \bibnamefont
  {Gruner}}, \bibinfo {author} {\bibfnamefont {O.}~\bibnamefont {Heczko}},
  \bibinfo {author} {\bibfnamefont {L.}~\bibnamefont {Schultz}}, \ and\
  \bibinfo {author} {\bibfnamefont {S.}~\bibnamefont {F{\"{a}}hler}},\ }\href
  {\doibase 10.1002/adem.201200058} {\bibfield  {journal} {\bibinfo  {journal}
  {Advanced Engineering Materials}\ }\textbf {\bibinfo {volume} {14}},\
  \bibinfo {pages} {562} (\bibinfo {year} {2012})}\BibitemShut {NoStop}%
\bibitem [{\citenamefont {Ullakko}\ \emph {et~al.}(1996)\citenamefont
  {Ullakko}, \citenamefont {Huang}, \citenamefont {Kantner}, \citenamefont
  {O'Handley},\ and\ \citenamefont {Kokorin}}]{Ullakko1996}%
  \BibitemOpen
  \bibfield  {author} {\bibinfo {author} {\bibfnamefont {K.}~\bibnamefont
  {Ullakko}}, \bibinfo {author} {\bibfnamefont {J.~K.}\ \bibnamefont {Huang}},
  \bibinfo {author} {\bibfnamefont {C.}~\bibnamefont {Kantner}}, \bibinfo
  {author} {\bibfnamefont {R.~C.}\ \bibnamefont {O'Handley}}, \ and\ \bibinfo
  {author} {\bibfnamefont {V.~V.}\ \bibnamefont {Kokorin}},\ }\href {\doibase
  10.1063/1.117637} {\bibfield  {journal} {\bibinfo  {journal} {Applied Physics
  Letters}\ }\textbf {\bibinfo {volume} {69}},\ \bibinfo {pages} {1966}
  (\bibinfo {year} {1996})}\BibitemShut {NoStop}%
\bibitem [{\citenamefont {Sozinov}\ \emph {et~al.}(2002)\citenamefont
  {Sozinov}, \citenamefont {Likhachev}, \citenamefont {Lanska},\ and\
  \citenamefont {Ullakko}}]{Sozinov2002}%
  \BibitemOpen
  \bibfield  {author} {\bibinfo {author} {\bibfnamefont {A.}~\bibnamefont
  {Sozinov}}, \bibinfo {author} {\bibfnamefont {A.~A.}\ \bibnamefont
  {Likhachev}}, \bibinfo {author} {\bibfnamefont {N.}~\bibnamefont {Lanska}}, \
  and\ \bibinfo {author} {\bibfnamefont {K.}~\bibnamefont {Ullakko}},\ }\href
  {\doibase 10.1063/1.1458075} {\bibfield  {journal} {\bibinfo  {journal}
  {Applied Physics Letters}\ }\textbf {\bibinfo {volume} {80}},\ \bibinfo
  {pages} {1746} (\bibinfo {year} {2002})}\BibitemShut {NoStop}%
\bibitem [{\citenamefont {Kainuma}\ \emph {et~al.}(2006)\citenamefont
  {Kainuma}, \citenamefont {Imano}, \citenamefont {Ito}, \citenamefont {Sutou},
  \citenamefont {Morito}, \citenamefont {Okamoto}, \citenamefont {Kitakami},
  \citenamefont {Oikawa}, \citenamefont {Fujita}, \citenamefont {Kanomata},\
  and\ \citenamefont {Ishida}}]{Kainuma2006}%
  \BibitemOpen
  \bibfield  {author} {\bibinfo {author} {\bibfnamefont {R.}~\bibnamefont
  {Kainuma}}, \bibinfo {author} {\bibfnamefont {Y.}~\bibnamefont {Imano}},
  \bibinfo {author} {\bibfnamefont {W.}~\bibnamefont {Ito}}, \bibinfo {author}
  {\bibfnamefont {Y.}~\bibnamefont {Sutou}}, \bibinfo {author} {\bibfnamefont
  {H.}~\bibnamefont {Morito}}, \bibinfo {author} {\bibfnamefont
  {S.}~\bibnamefont {Okamoto}}, \bibinfo {author} {\bibfnamefont
  {O.}~\bibnamefont {Kitakami}}, \bibinfo {author} {\bibfnamefont
  {K.}~\bibnamefont {Oikawa}}, \bibinfo {author} {\bibfnamefont
  {A.}~\bibnamefont {Fujita}}, \bibinfo {author} {\bibfnamefont
  {T.}~\bibnamefont {Kanomata}}, \ and\ \bibinfo {author} {\bibfnamefont
  {K.}~\bibnamefont {Ishida}},\ }\href {\doibase 10.1038/nature04493}
  {\bibfield  {journal} {\bibinfo  {journal} {Nature}\ }\textbf {\bibinfo
  {volume} {439}},\ \bibinfo {pages} {957} (\bibinfo {year}
  {2006})}\BibitemShut {NoStop}%
\bibitem [{\citenamefont {Planes}\ \emph {et~al.}(2009)\citenamefont {Planes},
  \citenamefont {Ma{\~{n}}osa},\ and\ \citenamefont {Acet}}]{Planes2009}%
  \BibitemOpen
  \bibfield  {author} {\bibinfo {author} {\bibfnamefont {A.}~\bibnamefont
  {Planes}}, \bibinfo {author} {\bibfnamefont {L.}~\bibnamefont
  {Ma{\~{n}}osa}}, \ and\ \bibinfo {author} {\bibfnamefont {M.}~\bibnamefont
  {Acet}},\ }\href {\doibase 10.1088/0953-8984/21/23/233201} {\bibfield
  {journal} {\bibinfo  {journal} {Journal of Physics: Condensed Matter}\
  }\textbf {\bibinfo {volume} {21}},\ \bibinfo {pages} {233201} (\bibinfo
  {year} {2009})}\BibitemShut {NoStop}%
\bibitem [{\citenamefont {Liu}\ \emph {et~al.}(2012)\citenamefont {Liu},
  \citenamefont {Gottschall}, \citenamefont {Skokov}, \citenamefont {Moore},\
  and\ \citenamefont {Gutfleisch}}]{Liu2012}%
  \BibitemOpen
  \bibfield  {author} {\bibinfo {author} {\bibfnamefont {J.}~\bibnamefont
  {Liu}}, \bibinfo {author} {\bibfnamefont {T.}~\bibnamefont {Gottschall}},
  \bibinfo {author} {\bibfnamefont {K.~P.}\ \bibnamefont {Skokov}}, \bibinfo
  {author} {\bibfnamefont {J.~D.}\ \bibnamefont {Moore}}, \ and\ \bibinfo
  {author} {\bibfnamefont {O.}~\bibnamefont {Gutfleisch}},\ }\href {\doibase
  10.1038/nmat3334} {\bibfield  {journal} {\bibinfo  {journal} {Nature
  materials}\ }\textbf {\bibinfo {volume} {11}},\ \bibinfo {pages} {620}
  (\bibinfo {year} {2012})}\BibitemShut {NoStop}%
\bibitem [{\citenamefont {Gruner}\ \emph {et~al.}(2018)\citenamefont {Gruner},
  \citenamefont {Niemann}, \citenamefont {Entel}, \citenamefont {Pentcheva},
  \citenamefont {R{\"{o}}{\ss}ler}, \citenamefont {Nielsch},\ and\
  \citenamefont {F{\"{a}}hler}}]{Gruner2018}%
  \BibitemOpen
  \bibfield  {author} {\bibinfo {author} {\bibfnamefont {M.~E.}\ \bibnamefont
  {Gruner}}, \bibinfo {author} {\bibfnamefont {R.}~\bibnamefont {Niemann}},
  \bibinfo {author} {\bibfnamefont {P.}~\bibnamefont {Entel}}, \bibinfo
  {author} {\bibfnamefont {R.}~\bibnamefont {Pentcheva}}, \bibinfo {author}
  {\bibfnamefont {U.~K.}\ \bibnamefont {R{\"{o}}{\ss}ler}}, \bibinfo {author}
  {\bibfnamefont {K.}~\bibnamefont {Nielsch}}, \ and\ \bibinfo {author}
  {\bibfnamefont {S.}~\bibnamefont {F{\"{a}}hler}},\ }\href {\doibase
  10.1038/s41598-018-26652-6} {\bibfield  {journal} {\bibinfo  {journal}
  {Scientific Reports}\ }\textbf {\bibinfo {volume} {8}},\ \bibinfo {pages}
  {8489} (\bibinfo {year} {2018})}\BibitemShut {NoStop}%
\bibitem [{\citenamefont {Fallot}(1938)}]{Fallot1938}%
  \BibitemOpen
  \bibfield  {author} {\bibinfo {author} {\bibfnamefont {M.}~\bibnamefont
  {Fallot}},\ }\href@noop {} {\bibfield  {journal} {\bibinfo  {journal} {Ann.
  Phys. (Paris)}\ }\textbf {\bibinfo {volume} {10}},\ \bibinfo {pages} {291}
  (\bibinfo {year} {1938})}\BibitemShut {NoStop}%
\bibitem [{\citenamefont {Fallot}\ and\ \citenamefont
  {Horcart}(1939)}]{Fallot1939}%
  \BibitemOpen
  \bibfield  {author} {\bibinfo {author} {\bibfnamefont {M.}~\bibnamefont
  {Fallot}}\ and\ \bibinfo {author} {\bibfnamefont {R.}~\bibnamefont
  {Horcart}},\ }\href@noop {} {\bibfield  {journal} {\bibinfo  {journal} {Rev.
  Sci.}\ }\textbf {\bibinfo {volume} {77}},\ \bibinfo {pages} {498}
  (\bibinfo {year} {1939})}\BibitemShut {NoStop}%
\bibitem [{\citenamefont {Wolloch}\ \emph {et~al.}(2016)\citenamefont
  {Wolloch}, \citenamefont {Gruner}, \citenamefont {Keune}, \citenamefont
  {Mohn}, \citenamefont {Redinger}, \citenamefont {Hofer}, \citenamefont
  {Suess}, \citenamefont {Podloucky}, \citenamefont {Landers}, \citenamefont
  {Salamon}, \citenamefont {Scheibel}, \citenamefont {Spoddig}, \citenamefont
  {Witte}, \citenamefont {{Roldan Cuenya}}, \citenamefont {Gutfleisch},
  \citenamefont {Hu}, \citenamefont {Zhao}, \citenamefont {Toellner},
  \citenamefont {Alp}, \citenamefont {Siewert}, \citenamefont {Entel},
  \citenamefont {Pentcheva},\ and\ \citenamefont {Wende}}]{Wolloch2016}%
  \BibitemOpen
  \bibfield  {author} {\bibinfo {author} {\bibfnamefont {M.}~\bibnamefont
  {Wolloch}}, \bibinfo {author} {\bibfnamefont {M.~E.}\ \bibnamefont {Gruner}},
  \bibinfo {author} {\bibfnamefont {W.}~\bibnamefont {Keune}}, \bibinfo
  {author} {\bibfnamefont {P.}~\bibnamefont {Mohn}}, \bibinfo {author}
  {\bibfnamefont {J.}~\bibnamefont {Redinger}}, \bibinfo {author}
  {\bibfnamefont {F.}~\bibnamefont {Hofer}}, \bibinfo {author} {\bibfnamefont
  {D.}~\bibnamefont {Suess}}, \bibinfo {author} {\bibfnamefont
  {R.}~\bibnamefont {Podloucky}}, \bibinfo {author} {\bibfnamefont
  {J.}~\bibnamefont {Landers}}, \bibinfo {author} {\bibfnamefont
  {S.}~\bibnamefont {Salamon}}, \bibinfo {author} {\bibfnamefont
  {F.}~\bibnamefont {Scheibel}}, \bibinfo {author} {\bibfnamefont
  {D.}~\bibnamefont {Spoddig}}, \bibinfo {author} {\bibfnamefont
  {R.}~\bibnamefont {Witte}}, \bibinfo {author} {\bibfnamefont
  {B.}~\bibnamefont {{Roldan Cuenya}}}, \bibinfo {author} {\bibfnamefont
  {O.}~\bibnamefont {Gutfleisch}}, \bibinfo {author} {\bibfnamefont {M.~Y.}\
  \bibnamefont {Hu}}, \bibinfo {author} {\bibfnamefont {J.}~\bibnamefont
  {Zhao}}, \bibinfo {author} {\bibfnamefont {T.}~\bibnamefont {Toellner}},
  \bibinfo {author} {\bibfnamefont {E.~E.}\ \bibnamefont {Alp}}, \bibinfo
  {author} {\bibfnamefont {M.}~\bibnamefont {Siewert}}, \bibinfo {author}
  {\bibfnamefont {P.}~\bibnamefont {Entel}}, \bibinfo {author} {\bibfnamefont
  {R.}~\bibnamefont {Pentcheva}}, \ and\ \bibinfo {author} {\bibfnamefont
  {H.}~\bibnamefont {Wende}},\ }\href {\doibase 10.1103/PhysRevB.94.174435}
  {\bibfield  {journal} {\bibinfo  {journal} {Physical Review B}\ }\textbf
  {\bibinfo {volume} {94}},\ \bibinfo {pages} {174435} (\bibinfo {year}
  {2016})},\ %\Eprint {http://arxiv.org/abs/1608.04268} {arXiv:1608.04268}
  \BibitemShut {NoStop}%
\bibitem [{\citenamefont {Lewis}\ \emph {et~al.}(2016)\citenamefont {Lewis},
  \citenamefont {Marrows},\ and\ \citenamefont {Langridge}}]{Lewis2016}%
  \BibitemOpen
  \bibfield  {author} {\bibinfo {author} {\bibfnamefont {L.~H.}\ \bibnamefont
  {Lewis}}, \bibinfo {author} {\bibfnamefont {C.~H.}\ \bibnamefont {Marrows}},
  \ and\ \bibinfo {author} {\bibfnamefont {S.}~\bibnamefont {Langridge}},\
  }\href {\doibase 10.1088/0022-3727/49/32/323002} {\bibfield  {journal}
  {\bibinfo  {journal} {Journal of Physics D: Applied Physics}\ }\textbf
  {\bibinfo {volume} {49}},\ \bibinfo {pages} {323002} (\bibinfo {year}
  {2016})}\BibitemShut {NoStop}%
\bibitem [{\citenamefont {Kim}\ \emph {et~al.}(2016)\citenamefont {Kim},
  \citenamefont {Ramesh},\ and\ \citenamefont {Kioussis}}]{Kim2016}%
  \BibitemOpen
  \bibfield  {author} {\bibinfo {author} {\bibfnamefont {J.}~\bibnamefont
  {Kim}}, \bibinfo {author} {\bibfnamefont {R.}~\bibnamefont {Ramesh}}, \ and\
  \bibinfo {author} {\bibfnamefont {N.}~\bibnamefont {Kioussis}},\ }\href
  {\doibase 10.1103/PhysRevB.94.180407} {\bibfield  {journal} {\bibinfo
  {journal} {Physical Review B}\ }\textbf {\bibinfo {volume} {94}},\ \bibinfo
  {pages} {180407(R)} (\bibinfo {year} {2016})}\BibitemShut {NoStop}%
\bibitem [{\citenamefont {Aschauer}\ \emph {et~al.}(2016)\citenamefont
  {Aschauer}, \citenamefont {Braddell}, \citenamefont {Brechb{\"{u}}hl},
  \citenamefont {Derlet},\ and\ \citenamefont {Spaldin}}]{Aschauer2016a}%
  \BibitemOpen
  \bibfield  {author} {\bibinfo {author} {\bibfnamefont {U.}~\bibnamefont
  {Aschauer}}, \bibinfo {author} {\bibfnamefont {R.}~\bibnamefont {Braddell}},
  \bibinfo {author} {\bibfnamefont {S.~A.}\ \bibnamefont {Brechb{\"{u}}hl}},
  \bibinfo {author} {\bibfnamefont {P.~M.}\ \bibnamefont {Derlet}}, \ and\
  \bibinfo {author} {\bibfnamefont {N.~A.}\ \bibnamefont {Spaldin}},\ }\href
  {\doibase 10.1103/PhysRevB.94.014109} {\bibfield  {journal} {\bibinfo
  {journal} {Physical Review B}\ }\textbf {\bibinfo {volume} {94}},\ \bibinfo
  {pages} {014109} (\bibinfo {year} {2016})},\ %\Eprint
  %{http://arxiv.org/abs/1603.01827v1} {arXiv:1603.01827v1} 
  \BibitemShut
  {NoStop}%
\bibitem [{\citenamefont {Bennett}\ \emph {et~al.}(2018)\citenamefont
  {Bennett}, \citenamefont {Herklotz}, \citenamefont {Cress}, \citenamefont
  {Ievlev}, \citenamefont {Rouleau}, \citenamefont {Mazin},\ and\ \citenamefont
  {Lauter}}]{Bennett2018}%
  \BibitemOpen
  \bibfield  {author} {\bibinfo {author} {\bibfnamefont {S.~P.}\ \bibnamefont
  {Bennett}}, \bibinfo {author} {\bibfnamefont {A.}~\bibnamefont {Herklotz}},
  \bibinfo {author} {\bibfnamefont {C.~D.}\ \bibnamefont {Cress}}, \bibinfo
  {author} {\bibfnamefont {A.}~\bibnamefont {Ievlev}}, \bibinfo {author}
  {\bibfnamefont {C.~M.}\ \bibnamefont {Rouleau}}, \bibinfo {author}
  {\bibfnamefont {I.~I.}\ \bibnamefont {Mazin}}, \ and\ \bibinfo {author}
  {\bibfnamefont {V.}~\bibnamefont {Lauter}},\ }\href {\doibase
  10.1080/21663831.2017.1402098} {\bibfield  {journal} {\bibinfo  {journal}
  {Materials Research Letters}\ }\textbf {\bibinfo {volume} {6}},\ \bibinfo
  {pages} {106} (\bibinfo {year} {2018})}\BibitemShut {NoStop}%
\bibitem [{\citenamefont {Gorji}\ \emph {et~al.}(2018)\citenamefont {Gorji},
  \citenamefont {Wang}, \citenamefont {Witte}, \citenamefont {Mu},
  \citenamefont {Kruk}, \citenamefont {K{\"{u}}bel},\ and\ \citenamefont
  {Hahn}}]{Gorji2018}%
  \BibitemOpen
  \bibfield  {author} {\bibinfo {author} {\bibfnamefont {M.~S.}\ \bibnamefont
  {Gorji}}, \bibinfo {author} {\bibfnamefont {D.}~\bibnamefont {Wang}},
  \bibinfo {author} {\bibfnamefont {R.}~\bibnamefont {Witte}}, \bibinfo
  {author} {\bibfnamefont {X.}~\bibnamefont {Mu}}, \bibinfo {author}
  {\bibfnamefont {R.}~\bibnamefont {Kruk}}, \bibinfo {author} {\bibfnamefont
  {C.}~\bibnamefont {K{\"{u}}bel}}, \ and\ \bibinfo {author} {\bibfnamefont
  {H.}~\bibnamefont {Hahn}},\ }\href {\doibase 10.1017/S1431927618005160}
  {\bibfield  {journal} {\bibinfo  {journal} {Microscopy and Microanalysis}\
  }\textbf {\bibinfo {volume} {24}},\ \bibinfo {pages} {934} (\bibinfo {year}
  {2018})}\BibitemShut {NoStop}%
\bibitem [{\citenamefont {Dr{\'{o}}{\.{z}}d{\.{z}}}\ \emph
  {et~al.}(2018{\natexlab{a}})\citenamefont {Dr{\'{o}}{\.{z}}d{\.{z}}},
  \citenamefont {{\'{S}}lȩzak}, \citenamefont {Matlak}, \citenamefont
  {Matlak}, \citenamefont {Freindl}, \citenamefont {Wilgocka-{\'{S}}lȩzak},
  \citenamefont {Spiridis}, \citenamefont {Korecki},\ and\ \citenamefont
  {{\'{S}}lȩzak}}]{Drozdz2018}%
  \BibitemOpen
  \bibfield  {author} {\bibinfo {author} {\bibfnamefont {P.}~\bibnamefont
  {Dr{\'{o}}{\.{z}}d{\.{z}}}}, \bibinfo {author} {\bibfnamefont
  {M.}~\bibnamefont {{\'{S}}l{\k{e}}zak}}, \bibinfo {author} {\bibfnamefont
  {K.}~\bibnamefont {Matlak}}, \bibinfo {author} {\bibfnamefont
  {B.}~\bibnamefont {Matlak}}, \bibinfo {author} {\bibfnamefont
  {K.}~\bibnamefont {Freindl}}, \bibinfo {author} {\bibfnamefont
  {D.}~\bibnamefont {Wilgocka-{\'{S}}l{\k{e}}zak}}, \bibinfo {author} {\bibfnamefont
  {N.}~\bibnamefont {Spiridis}}, \bibinfo {author} {\bibfnamefont
  {J.}~\bibnamefont {Korecki}}, \ and\ \bibinfo {author} {\bibfnamefont
  {T.}~\bibnamefont {{\'{S}}l{\k{e}}zak}},}\ \href {\doibase
  10.1103/PhysRevApplied.9.034030} {\bibfield  {journal} {\bibinfo  {journal}
  {Physical Review Applied}\ }\textbf {\bibinfo {volume} {9}},\ \bibinfo
  {pages} {034030} (\bibinfo {year} {2018}{\natexlab{a}})}\ \Eprint
  {} {} \BibitemShut {NoStop}%
\bibitem [{\citenamefont {Popescu}\ \emph {et~al.}(2018)\citenamefont
  {Popescu}, \citenamefont {Rodriguez-Lopez}, \citenamefont {Haney},\ and\
  \citenamefont {Woods}}]{Popescu2018}%
  \BibitemOpen
  \bibfield  {author} {\bibinfo {author} {\bibfnamefont {A.}~\bibnamefont
  {Popescu}}, \bibinfo {author} {\bibfnamefont {P.}~\bibnamefont
  {Rodriguez-Lopez}}, \bibinfo {author} {\bibfnamefont {P.~M.}\ \bibnamefont
  {Haney}}, \ and\ \bibinfo {author} {\bibfnamefont {L.~M.}\ \bibnamefont
  {Woods}},\ }\href {\doibase 10.1103/PhysRevB.97.140407} {\bibfield  {journal}
  {\bibinfo  {journal} {Physical Review B}\ }\textbf {\bibinfo {volume} {97}},\
  \bibinfo {pages} {140407} (\bibinfo {year} {2018})}\ \Eprint
  {} {} \BibitemShut {NoStop}%
\bibitem [{\citenamefont {Annaorazov}\ \emph {et~al.}(1996)\citenamefont
  {Annaorazov}, \citenamefont {Nikitin}, \citenamefont {Tyurin}, \citenamefont
  {Asatryan},\ and\ \citenamefont {Dovletov}}]{Annaorazov1996}%
  \BibitemOpen
  \bibfield  {author} {\bibinfo {author} {\bibfnamefont {M.~P.}\ \bibnamefont
  {Annaorazov}}, \bibinfo {author} {\bibfnamefont {S.~A.}\ \bibnamefont
  {Nikitin}}, \bibinfo {author} {\bibfnamefont {A.~L.}\ \bibnamefont {Tyurin}},
  \bibinfo {author} {\bibfnamefont {K.~A.}\ \bibnamefont {Asatryan}}, \ and\
  \bibinfo {author} {\bibfnamefont {A.~K.}\ \bibnamefont {Dovletov}},\ }\href
  {\doibase 10.1063/1.360955} {\bibfield  {journal} {\bibinfo  {journal}
  {Journal of Applied Physics}\ }\textbf {\bibinfo {volume} {79}},\ \bibinfo
  {pages} {1689} (\bibinfo {year} {1996})}\BibitemShut {NoStop}%
\bibitem [{\citenamefont {Chirkova}\ \emph {et~al.}(2016)\citenamefont
  {Chirkova}, \citenamefont {Skokov}, \citenamefont {Schultz}, \citenamefont
  {Baranov}, \citenamefont {Gutfleisch},\ and\ \citenamefont
  {Woodcock}}]{Chirkova2016}%
  \BibitemOpen
  \bibfield  {author} {\bibinfo {author} {\bibfnamefont {A.}~\bibnamefont
  {Chirkova}}, \bibinfo {author} {\bibfnamefont {K.}~\bibnamefont {Skokov}},
  \bibinfo {author} {\bibfnamefont {L.}~\bibnamefont {Schultz}}, \bibinfo
  {author} {\bibfnamefont {N.}~\bibnamefont {Baranov}}, \bibinfo {author}
  {\bibfnamefont {O.}~\bibnamefont {Gutfleisch}}, \ and\ \bibinfo {author}
  {\bibfnamefont {T.}~\bibnamefont {Woodcock}},\ }\href {\doibase
  10.1016/j.actamat.2015.11.054} {\bibfield  {journal} {\bibinfo  {journal}
  {Acta Materialia}\ }\textbf {\bibinfo {volume} {106}},\ \bibinfo {pages} {15}
  (\bibinfo {year} {2016})}\BibitemShut {NoStop}%
\bibitem [{\citenamefont {Stern-Taulats}\ \emph {et~al.}(2017)\citenamefont
  {Stern-Taulats}, \citenamefont {Cast{\'{a}}n}, \citenamefont {Planes},
  \citenamefont {Lewis}, \citenamefont {Barua}, \citenamefont {Pramanick},
  \citenamefont {Majumdar},\ and\ \citenamefont
  {Ma{\~{n}}osa}}]{Stern-Taulats2017}%
  \BibitemOpen
  \bibfield  {author} {\bibinfo {author} {\bibfnamefont {E.}~\bibnamefont
  {Stern-Taulats}}, \bibinfo {author} {\bibfnamefont {T.}~\bibnamefont
  {Cast{\'{a}}n}}, \bibinfo {author} {\bibfnamefont {A.}~\bibnamefont
  {Planes}}, \bibinfo {author} {\bibfnamefont {L.~H.}\ \bibnamefont {Lewis}},
  \bibinfo {author} {\bibfnamefont {R.}~\bibnamefont {Barua}}, \bibinfo
  {author} {\bibfnamefont {S.}~\bibnamefont {Pramanick}}, \bibinfo {author}
  {\bibfnamefont {S.}~\bibnamefont {Majumdar}}, \ and\ \bibinfo {author}
  {\bibfnamefont {L.}~\bibnamefont {Ma{\~{n}}osa}},\ }\href {\doibase
  10.1103/PhysRevB.95.104424} {\bibfield  {journal} {\bibinfo  {journal}
  {Physical Review B}\ }\textbf {\bibinfo {volume} {95}},\ \bibinfo {pages}
  {104424} (\bibinfo {year} {2017})}\BibitemShut {NoStop}%
\bibitem [{\citenamefont {Thiele}\ \emph {et~al.}(2003)\citenamefont {Thiele},
  \citenamefont {Maat},\ and\ \citenamefont {Fullerton}}]{Thiele2003a}%
  \BibitemOpen
  \bibfield  {author} {\bibinfo {author} {\bibfnamefont {J.-U.}\ \bibnamefont
  {Thiele}}, \bibinfo {author} {\bibfnamefont {S.}~\bibnamefont {Maat}}, \ and\
  \bibinfo {author} {\bibfnamefont {E.~E.}\ \bibnamefont {Fullerton}},\ }\href
  {\doibase 10.1063/1.1571232} {\bibfield  {journal} {\bibinfo  {journal}
  {Applied Physics Letters}\ }\textbf {\bibinfo {volume} {82}},\ \bibinfo
  {pages} {2859} (\bibinfo {year} {2003})}\BibitemShut {NoStop}%
\bibitem [{\citenamefont {Bordel}\ \emph {et~al.}(2012)\citenamefont {Bordel},
  \citenamefont {Juraszek}, \citenamefont {Cooke}, \citenamefont
  {Baldasseroni}, \citenamefont {Mankovsky}, \citenamefont {Min{\'{a}}r},
  \citenamefont {Ebert}, \citenamefont {Moyerman}, \citenamefont {Fullerton},\
  and\ \citenamefont {Hellman}}]{Bordel2012}%
  \BibitemOpen
  \bibfield  {author} {\bibinfo {author} {\bibfnamefont {C.}~\bibnamefont
  {Bordel}}, \bibinfo {author} {\bibfnamefont {J.}~\bibnamefont {Juraszek}},
  \bibinfo {author} {\bibfnamefont {D.~W.}\ \bibnamefont {Cooke}}, \bibinfo
  {author} {\bibfnamefont {C.}~\bibnamefont {Baldasseroni}}, \bibinfo {author}
  {\bibfnamefont {S.}~\bibnamefont {Mankovsky}}, \bibinfo {author}
  {\bibfnamefont {J.}~\bibnamefont {Min{\'{a}}r}}, \bibinfo {author}
  {\bibfnamefont {H.}~\bibnamefont {Ebert}}, \bibinfo {author} {\bibfnamefont
  {S.}~\bibnamefont {Moyerman}}, \bibinfo {author} {\bibfnamefont {E.~E.}\
  \bibnamefont {Fullerton}}, \ and\ \bibinfo {author} {\bibfnamefont
  {F.}~\bibnamefont {Hellman}},\ }\href {\doibase
  10.1103/PhysRevLett.109.117201} {\bibfield  {journal} {\bibinfo  {journal}
  {Physical Review Letters}\ }\textbf {\bibinfo {volume} {109}},\ \bibinfo
  {pages} {117201} (\bibinfo {year} {2012})}\BibitemShut {NoStop}%
\bibitem [{\citenamefont {Marti}\ \emph {et~al.}(2014)\citenamefont {Marti},
  \citenamefont {Fina}, \citenamefont {Frontera}, \citenamefont {Liu},
  \citenamefont {Wadley}, \citenamefont {He}, \citenamefont {Paull},
  \citenamefont {Clarkson}, \citenamefont {Kudrnovsk{\'{y}}}, \citenamefont
  {Turek}, \citenamefont {Kune{\v{s}}}, \citenamefont {Yi}, \citenamefont
  {Chu}, \citenamefont {Nelson}, \citenamefont {You}, \citenamefont {Arenholz},
  \citenamefont {Salahuddin}, \citenamefont {Fontcuberta}, \citenamefont
  {Jungwirth},\ and\ \citenamefont {Ramesh}}]{Marti2014}%
  \BibitemOpen
  \bibfield  {author} {\bibinfo {author} {\bibfnamefont {X.}~\bibnamefont
  {Marti}}, \bibinfo {author} {\bibfnamefont {I.}~\bibnamefont {Fina}},
  \bibinfo {author} {\bibfnamefont {C.}~\bibnamefont {Frontera}}, \bibinfo
  {author} {\bibfnamefont {J.}~\bibnamefont {Liu}}, \bibinfo {author}
  {\bibfnamefont {P.}~\bibnamefont {Wadley}}, \bibinfo {author} {\bibfnamefont
  {Q.}~\bibnamefont {He}}, \bibinfo {author} {\bibfnamefont {R.~J.}\
  \bibnamefont {Paull}}, \bibinfo {author} {\bibfnamefont {J.~D.}\ \bibnamefont
  {Clarkson}}, \bibinfo {author} {\bibfnamefont {J.}~\bibnamefont
  {Kudrnovsk{\'{y}}}}, \bibinfo {author} {\bibfnamefont {I.}~\bibnamefont
  {Turek}}, \bibinfo {author} {\bibfnamefont {J.}~\bibnamefont {Kune{\v{s}}}},
  \bibinfo {author} {\bibfnamefont {D.}~\bibnamefont {Yi}}, \bibinfo {author}
  {\bibfnamefont {J.-H.}\ \bibnamefont {Chu}}, \bibinfo {author} {\bibfnamefont
  {C.~T.}\ \bibnamefont {Nelson}}, \bibinfo {author} {\bibfnamefont
  {L.}~\bibnamefont {You}}, \bibinfo {author} {\bibfnamefont {E.}~\bibnamefont
  {Arenholz}}, \bibinfo {author} {\bibfnamefont {S.}~\bibnamefont
  {Salahuddin}}, \bibinfo {author} {\bibfnamefont {J.}~\bibnamefont
  {Fontcuberta}}, \bibinfo {author} {\bibfnamefont {T.}~\bibnamefont
  {Jungwirth}}, \ and\ \bibinfo {author} {\bibfnamefont {R.}~\bibnamefont
  {Ramesh}},\ }\href {\doibase 10.1038/nmat3861} {\bibfield  {journal}
  {\bibinfo  {journal} {Nature Materials}\ }\textbf {\bibinfo {volume} {13}},\
  \bibinfo {pages} {367} (\bibinfo {year} {2014})}\BibitemShut {NoStop}%
\bibitem [{\citenamefont {Dr{\'{o}}{\.{z}}d{\.{z}}}\ \emph
  {et~al.}(2018{\natexlab{b}})\citenamefont {Dr{\'{o}}{\.{z}}d{\.{z}}},
  \citenamefont {{\'{S}}l{\c{e}}zak}, \citenamefont {Matlak}, \citenamefont
  {Kozio{\l}-Rachwa{\l}}, \citenamefont {Wilgocka-{\'{S}}l{\c{e}}zak},
  \citenamefont {Korecki},\ and\ \citenamefont
  {{\'{S}}l{\c{e}}zak}}]{Drozdz2018a}%
  \BibitemOpen
  \bibfield  {author} {\bibinfo {author} {\bibfnamefont {P.}~\bibnamefont
  {Dr{\'{o}}{\.{z}}d{\.{z}}}}, \bibinfo {author} {\bibfnamefont
  {M.}~\bibnamefont {{\'{S}}l{\c{e}}zak}}, \bibinfo {author} {\bibfnamefont
  {K.}~\bibnamefont {Matlak}}, \bibinfo {author} {\bibfnamefont
  {A.}~\bibnamefont {Kozio{\l}-Rachwa{\l}}}, \bibinfo {author} {\bibfnamefont
  {D.}~\bibnamefont {Wilgocka-{\'{S}}l{\c{e}}zak}}, \bibinfo {author}
  {\bibfnamefont {J.}~\bibnamefont {Korecki}}, \ and\ \bibinfo {author}
  {\bibfnamefont {T.}~\bibnamefont {{\'{S}}l{\c{e}}zak}},\ }\href {\doibase
  10.1063/1.5042841} {\bibfield  {journal} {\bibinfo  {journal} {AIP Advances}\
  }\textbf {\bibinfo {volume} {8}},\ \bibinfo {pages} {101434} (\bibinfo {year}
  {2018}{\natexlab{b}})}\BibitemShut {NoStop}%
\bibitem [{\citenamefont {Witte}\ \emph {et~al.}(2016)\citenamefont {Witte},
  \citenamefont {Kruk}, \citenamefont {Gruner}, \citenamefont {Brand},
  \citenamefont {Wang}, \citenamefont {Schlabach}, \citenamefont {Beck},
  \citenamefont {Provenzano}, \citenamefont {Pentcheva}, \citenamefont
  {Wende},\ and\ \citenamefont {Hahn}}]{Witte2016}%
  \BibitemOpen
  \bibfield  {author} {\bibinfo {author} {\bibfnamefont {R.}~\bibnamefont
  {Witte}}, \bibinfo {author} {\bibfnamefont {R.}~\bibnamefont {Kruk}},
  \bibinfo {author} {\bibfnamefont {M.~E.}\ \bibnamefont {Gruner}}, \bibinfo
  {author} {\bibfnamefont {R.~A.}\ \bibnamefont {Brand}}, \bibinfo {author}
  {\bibfnamefont {D.}~\bibnamefont {Wang}}, \bibinfo {author} {\bibfnamefont
  {S.}~\bibnamefont {Schlabach}}, \bibinfo {author} {\bibfnamefont
  {A.}~\bibnamefont {Beck}}, \bibinfo {author} {\bibfnamefont {V.}~\bibnamefont
  {Provenzano}}, \bibinfo {author} {\bibfnamefont {R.}~\bibnamefont
  {Pentcheva}}, \bibinfo {author} {\bibfnamefont {H.}~\bibnamefont {Wende}}, \
  and\ \bibinfo {author} {\bibfnamefont {H.}~\bibnamefont {Hahn}},\ }\href
  {\doibase 10.1103/PhysRevB.93.104416} {\bibfield  {journal} {\bibinfo
  {journal} {Physical Review B}\ }\textbf {\bibinfo {volume} {93}},\ \bibinfo
  {pages} {104416} (\bibinfo {year} {2016})}\BibitemShut {NoStop}%
\bibitem[]{SM} See Supplemental Material at [URL will be
  inserted by publisher] for additional data, reflection high energy electron diffraction patterns and conversion electron M\"ossbauer spectra, as
well as a graphical illustration of the computational procedures.
 
\bibitem [{\citenamefont {Uebayashi}\ \emph {et~al.}(2006)\citenamefont
  {Uebayashi}, \citenamefont {Shimizu},\ and\ \citenamefont
  {Yamada}}]{Uebayashi2006}%
  \BibitemOpen
  \bibfield  {author} {\bibinfo {author} {\bibfnamefont {K.}~\bibnamefont
  {Uebayashi}}, \bibinfo {author} {\bibfnamefont {H.}~\bibnamefont {Shimizu}},
  \ and\ \bibinfo {author} {\bibfnamefont {H.}~\bibnamefont {Yamada}},\ }\href
  {\doibase 10.2320/matertrans.47.456} {\bibfield  {journal} {\bibinfo
  {journal} {Materials transactions}\ }\textbf {\bibinfo {volume} {47}},\
  \bibinfo {pages} {456} (\bibinfo {year} {2006})}\BibitemShut {NoStop}%
\bibitem [{\citenamefont {Staunton}\ \emph {et~al.}(2014)\citenamefont
  {Staunton}, \citenamefont {Banerjee}, \citenamefont {{dos Santos Dias}},
  \citenamefont {Deak},\ and\ \citenamefont {Szunyogh}}]{Staunton2014}%
  \BibitemOpen
  \bibfield  {author} {\bibinfo {author} {\bibfnamefont {J.~B.}\ \bibnamefont
  {Staunton}}, \bibinfo {author} {\bibfnamefont {R.}~\bibnamefont {Banerjee}},
  \bibinfo {author} {\bibfnamefont {M.}~\bibnamefont {{dos Santos Dias}}},
  \bibinfo {author} {\bibfnamefont {A.}~\bibnamefont {Deak}}, \ and\ \bibinfo
  {author} {\bibfnamefont {L.}~\bibnamefont {Szunyogh}},\ }\href {\doibase
  10.1103/PhysRevB.89.054427} {\bibfield  {journal} {\bibinfo  {journal}
  {Physical Review B}\ }\textbf {\bibinfo {volume} {89}},\ \bibinfo {pages}
  {054427} (\bibinfo {year} {2014})}\BibitemShut {NoStop}%
\bibitem [{\citenamefont {Kresse}\ and\ \citenamefont
  {Furthm{\"{u}}ller}(1996)}]{VASP1}%
  \BibitemOpen
  \bibfield  {author} {\bibinfo {author} {\bibfnamefont {G.}~\bibnamefont
  {Kresse}}\ and\ \bibinfo {author} {\bibfnamefont {J.}~\bibnamefont
  {Furthm{\"{u}}ller}},\ }\href {\doibase 10.1103/PhysRevB.54.11169} {\bibfield
   {journal} {\bibinfo  {journal} {Physical Review B}\ }\textbf {\bibinfo
  {volume} {54}},\ \bibinfo {pages} {11169} (\bibinfo {year}
  {1996})}\BibitemShut {NoStop}%
\bibitem [{\citenamefont {Kresse}\ and\ \citenamefont {Joubert}(1999)}]{VASP2}%
  \BibitemOpen
  \bibfield  {author} {\bibinfo {author} {\bibfnamefont {G.}~\bibnamefont
  {Kresse}}\ and\ \bibinfo {author} {\bibfnamefont {D.}~\bibnamefont
  {Joubert}},\ }\href {\doibase 10.1103/PhysRevB.59.1758} {\bibfield  {journal}
  {\bibinfo  {journal} {Physical Review B}\ }\textbf {\bibinfo {volume} {59}},\
  \bibinfo {pages} {1758} (\bibinfo {year} {1999})}\BibitemShut {NoStop}%
\bibitem [{\citenamefont {Perdew}\ \emph {et~al.}(1996)\citenamefont {Perdew},
  \citenamefont {Burke},\ and\ \citenamefont {Ernzerhoff}}]{cn:Perdew96}%
  \BibitemOpen
  \bibfield  {author} {\bibinfo {author} {\bibfnamefont {J.~P.}\ \bibnamefont
  {Perdew}}, \bibinfo {author} {\bibfnamefont {K.}~\bibnamefont {Burke}}, \
  and\ \bibinfo {author} {\bibfnamefont {M.}~\bibnamefont {Ernzerhoff}},\
  }\href@noop {} {\bibfield  {journal} {\bibinfo  {journal} {Phys. Rev. Lett.}\
  }\textbf {\bibinfo {volume} {77}},\ \bibinfo {pages} {3865} (\bibinfo {year}
  {1996})}\BibitemShut {NoStop}%
\bibitem [{\citenamefont {Zunger}\ \emph {et~al.}(1990)\citenamefont {Zunger},
  \citenamefont {Wei}, \citenamefont {Ferreira},\ and\ \citenamefont
  {Bernard}}]{Zunger1990}%
  \BibitemOpen
  \bibfield  {author} {\bibinfo {author} {\bibfnamefont {A.}~\bibnamefont
  {Zunger}}, \bibinfo {author} {\bibfnamefont {S.-H.}\ \bibnamefont {Wei}},
  \bibinfo {author} {\bibfnamefont {L.~G.}\ \bibnamefont {Ferreira}}, \ and\
  \bibinfo {author} {\bibfnamefont {J.~E.}\ \bibnamefont {Bernard}},\ }\href
  {\doibase 10.1103/PhysRevLett.65.353} {\bibfield  {journal} {\bibinfo
  {journal} {Physical Review Letters}\ }\textbf {\bibinfo {volume} {65}},\
  \bibinfo {pages} {353} (\bibinfo {year} {1990})}\BibitemShut {NoStop}%
\bibitem [{\citenamefont {Jiang}(2009)}]{cn:Jiang09Acta}%
  \BibitemOpen
  \bibfield  {author} {\bibinfo {author} {\bibfnamefont {C.}~\bibnamefont
  {Jiang}},\ }\href@noop {} {\bibfield  {journal} {\bibinfo  {journal} {Acta
  Mater.}\ }\textbf {\bibinfo {volume} {57}},\ \bibinfo {pages} {4716}
  (\bibinfo {year} {2009})}\BibitemShut {NoStop}%
\bibitem [{\citenamefont {Methfessel}\ and\ \citenamefont
  {Paxton}(1989)}]{cn:Methfessel89}%
  \BibitemOpen
  \bibfield  {author} {\bibinfo {author} {\bibfnamefont {M.}~\bibnamefont
  {Methfessel}}\ and\ \bibinfo {author} {\bibfnamefont {A.~T.}\ \bibnamefont
  {Paxton}},\ }\href {\doibase 10.1103/PhysRevB.40.3616} {\bibfield  {journal}
  {\bibinfo  {journal} {Phys. Rev. B}\ }\textbf {\bibinfo {volume} {40}},\
  \bibinfo {pages} {3616} (\bibinfo {year} {1989})}\BibitemShut {NoStop}%
\bibitem [{\citenamefont {Bl\"ochl}\ \emph {et~al.}(1994)\citenamefont
  {Bl\"ochl}, \citenamefont {Jepsen},\ and\ \citenamefont
  {Andersen}}]{cn:Bloechl94}%
  \BibitemOpen
  \bibfield  {author} {\bibinfo {author} {\bibfnamefont {P.~E.}\ \bibnamefont
  {Bl\"ochl}}, \bibinfo {author} {\bibfnamefont {O.}~\bibnamefont {Jepsen}}, \
  and\ \bibinfo {author} {\bibfnamefont {O.~K.}\ \bibnamefont {Andersen}},\
  }\href {\doibase 10.1103/PhysRevB.49.16223} {\bibfield  {journal} {\bibinfo
  {journal} {Phys. Rev. B}\ }\textbf {\bibinfo {volume} {49}},\ \bibinfo
  {pages} {16223} (\bibinfo {year} {1994})}\BibitemShut {NoStop}%
\bibitem [{\citenamefont {Featherston}\ and\ \citenamefont
  {Neighbours}(1963)}]{Featherston1963}%
  \BibitemOpen
  \bibfield  {author} {\bibinfo {author} {\bibfnamefont {F.~H.}\ \bibnamefont
  {Featherston}}\ and\ \bibinfo {author} {\bibfnamefont {J.~R.}\ \bibnamefont
  {Neighbours}},\ }\href {\doibase 10.1103/PhysRev.130.1324} {\bibfield
  {journal} {\bibinfo  {journal} {Physical Review}\ }\textbf {\bibinfo {volume}
  {130}},\ \bibinfo {pages} {1324} (\bibinfo {year} {1963})}\BibitemShut
  {NoStop}%
\bibitem [{\citenamefont {Alers}(1960)}]{Alers1960}%
  \BibitemOpen
  \bibfield  {author} {\bibinfo {author} {\bibfnamefont {G.~A.}\ \bibnamefont
  {Alers}},\ }\href {\doibase 10.1103/PhysRev.119.1532} {\bibfield  {journal}
  {\bibinfo  {journal} {Physical Review}\ }\textbf {\bibinfo {volume} {119}},\
  \bibinfo {pages} {1532} (\bibinfo {year} {1960})}\BibitemShut {NoStop}%
\bibitem [{\citenamefont {Predel}(1998)}]{Predel1998}%
  \BibitemOpen
  \bibfield  {author} {\bibinfo {author} {\bibfnamefont {B.}~\bibnamefont
  {Predel}},\ }\href {\doibase 10.1007/10551312_2869} {\emph {\bibinfo {title}
  {Pu-Re - Zn-Zr}}},\ edited by\ \bibinfo {editor} {\bibfnamefont
  {O.}~\bibnamefont {Madelung}},\ Landolt-B{\"{o}}rnstein - Group IV Physical
  Chemistry\ (\bibinfo  {publisher} {Springer-Verlag},\ \bibinfo {address}
  {Berlin/Heidelberg},\ \bibinfo {year} {1998})\ pp.\ \bibinfo {pages}
  {1--3}\BibitemShut {NoStop}%
\bibitem [{\citenamefont {Swartzendruber}\ and\ \citenamefont
  {Sundman}(1983)}]{Swartzendruber1983}%
  \BibitemOpen
  \bibfield  {author} {\bibinfo {author} {\bibfnamefont {L.~J.}\ \bibnamefont
  {Swartzendruber}}\ and\ \bibinfo {author} {\bibfnamefont {B.}~\bibnamefont
  {Sundman}},\ }\href {\doibase 10.1007/BF02884862} {\bibfield  {journal}
  {\bibinfo  {journal} {Bulletin of Alloy Phase Diagrams}\ }\textbf {\bibinfo
  {volume} {4}},\ \bibinfo {pages} {155} (\bibinfo {year} {1983})}\BibitemShut
  {NoStop}%
\bibitem [{\citenamefont {Spie{\ss}}\ \emph {et~al.}(2009)\citenamefont
  {Spie{\ss}}, \citenamefont {Teichert}, \citenamefont {Schwarzer},
  \citenamefont {Behnken},\ and\ \citenamefont {Genzel}}]{Spiess2009}%
  \BibitemOpen
  \bibfield  {author} {\bibinfo {author} {\bibfnamefont {L.}~\bibnamefont
  {Spie{\ss}}}, \bibinfo {author} {\bibfnamefont {G.}~\bibnamefont {Teichert}},
  \bibinfo {author} {\bibfnamefont {R.}~\bibnamefont {Schwarzer}}, \bibinfo
  {author} {\bibfnamefont {H.}~\bibnamefont {Behnken}}, \ and\ \bibinfo
  {author} {\bibfnamefont {C.}~\bibnamefont {Genzel}},\ }\href {\doibase
  10.1007/978-3-8349-9434-9} {\emph {\bibinfo {title} {Moderne
  R{\"{o}}ntgenbeugung}}}\ (\bibinfo  {publisher} {Vieweg+Teubner},\ \bibinfo
  {address} {Wiesbaden},\ \bibinfo {year} {2009})\ pp.\ \bibinfo {pages}
  {5--40}\BibitemShut {NoStop}%
\bibitem [{\citenamefont {H{\"{y}}tch}\ \emph {et~al.}(1998)\citenamefont
  {H{\"{y}}tch}, \citenamefont {Snoeck},\ and\ \citenamefont
  {Kilaas}}]{Hytch1998}%
  \BibitemOpen
  \bibfield  {author} {\bibinfo {author} {\bibfnamefont {M.}~\bibnamefont
  {H{\"{y}}tch}}, \bibinfo {author} {\bibfnamefont {E.}~\bibnamefont {Snoeck}},
  \ and\ \bibinfo {author} {\bibfnamefont {R.}~\bibnamefont {Kilaas}},\ }\href
  {\doibase 10.1016/S0304-3991(98)00035-7} {\bibfield  {journal} {\bibinfo
  {journal} {Ultramicroscopy}\ }\textbf {\bibinfo {volume} {74}},\ \bibinfo
  {pages} {131} (\bibinfo {year} {1998})}\BibitemShut {NoStop}%
\bibitem [{\citenamefont {Witte}\ \emph {et~al.}(2017)\citenamefont {Witte},
  \citenamefont {Kruk}, \citenamefont {Molinari}, \citenamefont {Wang},
  \citenamefont {Schlabach}, \citenamefont {Brand}, \citenamefont
  {Provenzano},\ and\ \citenamefont {Hahn}}]{Witte2017}%
  \BibitemOpen
  \bibfield  {author} {\bibinfo {author} {\bibfnamefont {R.}~\bibnamefont
  {Witte}}, \bibinfo {author} {\bibfnamefont {R.}~\bibnamefont {Kruk}},
  \bibinfo {author} {\bibfnamefont {A.}~\bibnamefont {Molinari}}, \bibinfo
  {author} {\bibfnamefont {D.}~\bibnamefont {Wang}}, \bibinfo {author}
  {\bibfnamefont {S.}~\bibnamefont {Schlabach}}, \bibinfo {author}
  {\bibfnamefont {R.~A.}\ \bibnamefont {Brand}}, \bibinfo {author}
  {\bibfnamefont {V.}~\bibnamefont {Provenzano}}, \ and\ \bibinfo {author}
  {\bibfnamefont {H.}~\bibnamefont {Hahn}},\ }\href {\doibase
  10.1088/1361-6463/50/2/025007} {\bibfield  {journal} {\bibinfo  {journal}
  {Journal of Physics D: Applied Physics}\ }\textbf {\bibinfo {volume} {50}},\
  \bibinfo {pages} {025007} (\bibinfo {year} {2017})}\BibitemShut {NoStop}%
\bibitem [{\citenamefont {Opahle}\ \emph {et~al.}(2009)\citenamefont {Opahle},
  \citenamefont {Koepernik}, \citenamefont {Nitzsche},\ and\ \citenamefont
  {Richter}}]{cn:Opahle09}%
  \BibitemOpen
  \bibfield  {author} {\bibinfo {author} {\bibfnamefont {I.}~\bibnamefont
  {Opahle}}, \bibinfo {author} {\bibfnamefont {K.}~\bibnamefont {Koepernik}},
  \bibinfo {author} {\bibfnamefont {U.}~\bibnamefont {Nitzsche}}, \ and\
  \bibinfo {author} {\bibfnamefont {M.}~\bibnamefont {Richter}},\ }\href@noop
  {} {\bibfield  {journal} {\bibinfo  {journal} {Appl. Phys. Lett.}\ }\textbf
  {\bibinfo {volume} {94}},\ \bibinfo {pages} {072508} (\bibinfo {year}
  {2009})}\BibitemShut {NoStop}%
\bibitem [{\citenamefont {Kauffmann-Weiss}\ \emph {et~al.}(2011)\citenamefont
  {Kauffmann-Weiss}, \citenamefont {Gruner}, \citenamefont {Backen},
  \citenamefont {Schultz}, \citenamefont {Entel},\ and\ \citenamefont
  {F\"ahler}}]{cn:Weiss11PRL}%
  \BibitemOpen
  \bibfield  {author} {\bibinfo {author} {\bibfnamefont {S.}~\bibnamefont
  {Kauffmann-Weiss}}, \bibinfo {author} {\bibfnamefont {M.~E.}\ \bibnamefont
  {Gruner}}, \bibinfo {author} {\bibfnamefont {A.}~\bibnamefont {Backen}},
  \bibinfo {author} {\bibfnamefont {L.}~\bibnamefont {Schultz}}, \bibinfo
  {author} {\bibfnamefont {P.}~\bibnamefont {Entel}}, \ and\ \bibinfo {author}
  {\bibfnamefont {S.}~\bibnamefont {F\"ahler}},\ }\href@noop {} {\bibfield
  {journal} {\bibinfo  {journal} {Phys. Rev. Lett.}\ }\textbf {\bibinfo
  {volume} {107}},\ \bibinfo {pages} {206105} (\bibinfo {year}
  {2011})}\BibitemShut {NoStop}%
\bibitem [{\citenamefont {Gruner}\ and\ \citenamefont
  {Entel}(2011)}]{cn:Gruner11PRB}%
  \BibitemOpen
  \bibfield  {author} {\bibinfo {author} {\bibfnamefont {M.~E.}\ \bibnamefont
  {Gruner}}\ and\ \bibinfo {author} {\bibfnamefont {P.}~\bibnamefont {Entel}},\
  }\href@noop {} {\bibfield  {journal} {\bibinfo  {journal} {Phys. Rev. B}\
  }\textbf {\bibinfo {volume} {83}},\ \bibinfo {pages} {214415} (\bibinfo
  {year} {2011})}\BibitemShut {NoStop}%
\bibitem [{\citenamefont {Ayers}\ \emph {et~al.}(1991)\citenamefont {Ayers},
  \citenamefont {Ghandhi},\ and\ \citenamefont {Schowalter}}]{Ayers1991}%
  \BibitemOpen
  \bibfield  {author} {\bibinfo {author} {\bibfnamefont {J.}~\bibnamefont
  {Ayers}}, \bibinfo {author} {\bibfnamefont {S.}~\bibnamefont {Ghandhi}}, \
  and\ \bibinfo {author} {\bibfnamefont {L.}~\bibnamefont {Schowalter}},\
  }\href {\doibase 10.1016/0022-0248(91)90077-I} {\bibfield  {journal}
  {\bibinfo  {journal} {Journal of Crystal Growth}\ }\textbf {\bibinfo {volume}
  {113}},\ \bibinfo {pages} {430} (\bibinfo {year} {1991})}\BibitemShut
  {NoStop}%
\bibitem [{\citenamefont {Kim}\ \emph {et~al.}(2006)\citenamefont {Kim},
  \citenamefont {Morioka}, \citenamefont {Ueno}, \citenamefont {Yokoyama},
  \citenamefont {Funakubo}, \citenamefont {Lee},\ and\ \citenamefont
  {Baik}}]{Kim2006}%
  \BibitemOpen
  \bibfield  {author} {\bibinfo {author} {\bibfnamefont {Y.~K.}\ \bibnamefont
  {Kim}}, \bibinfo {author} {\bibfnamefont {H.}~\bibnamefont {Morioka}},
  \bibinfo {author} {\bibfnamefont {R.}~\bibnamefont {Ueno}}, \bibinfo {author}
  {\bibfnamefont {S.}~\bibnamefont {Yokoyama}}, \bibinfo {author}
  {\bibfnamefont {H.}~\bibnamefont {Funakubo}}, \bibinfo {author}
  {\bibfnamefont {K.}~\bibnamefont {Lee}}, \ and\ \bibinfo {author}
  {\bibfnamefont {S.}~\bibnamefont {Baik}},\ }\href {\doibase
  10.1063/1.2214169} {\bibfield  {journal} {\bibinfo  {journal} {Applied
  Physics Letters}\ }\textbf {\bibinfo {volume} {88}},\ \bibinfo {pages}
  {252904} (\bibinfo {year} {2006})}\BibitemShut {NoStop}%
\bibitem [{\citenamefont {Lee}\ and\ \citenamefont {Baik}(2000)}]{Lee2000b}%
  \BibitemOpen
  \bibfield  {author} {\bibinfo {author} {\bibfnamefont {K.~S.}\ \bibnamefont
  {Lee}}\ and\ \bibinfo {author} {\bibfnamefont {S.}~\bibnamefont {Baik}},\
  }\href {\doibase 10.1063/1.373493} {\bibfield  {journal} {\bibinfo  {journal}
  {Journal of Applied Physics}\ }\textbf {\bibinfo {volume} {87}},\ \bibinfo
  {pages} {8035} (\bibinfo {year} {2000})}\BibitemShut {NoStop}%
\bibitem [{\citenamefont {Schowalter}\ \emph {et~al.}(1990)\citenamefont
  {Schowalter}, \citenamefont {Hall}, \citenamefont {Lewis},\ and\
  \citenamefont {{Shin Hashimoto}}}]{Schowalter1990}%
  \BibitemOpen
  \bibfield  {author} {\bibinfo {author} {\bibfnamefont {L.}~\bibnamefont
  {Schowalter}}, \bibinfo {author} {\bibfnamefont {E.}~\bibnamefont {Hall}},
  \bibinfo {author} {\bibfnamefont {N.}~\bibnamefont {Lewis}}, \ and\ \bibinfo
  {author} {\bibnamefont {{Shin Hashimoto}}},\ }\href {\doibase
  10.1016/0040-6090(90)90442-G} {\bibfield  {journal} {\bibinfo  {journal}
  {Thin Solid Films}\ }\textbf {\bibinfo {volume} {184}},\ \bibinfo {pages}
  {437} (\bibinfo {year} {1990})}\BibitemShut {NoStop}%
\bibitem [{\citenamefont {Ohnuma}\ \emph {et~al.}(2009)\citenamefont {Ohnuma},
  \citenamefont {Gendo}, \citenamefont {Kainuma}, \citenamefont {Inden},\ and\
  \citenamefont {Ishida}}]{Ohnuma2009}%
  \BibitemOpen
  \bibfield  {author} {\bibinfo {author} {\bibfnamefont {I.}~\bibnamefont
  {Ohnuma}}, \bibinfo {author} {\bibfnamefont {T.}~\bibnamefont {Gendo}},
  \bibinfo {author} {\bibfnamefont {R.}~\bibnamefont {Kainuma}}, \bibinfo
  {author} {\bibfnamefont {G.}~\bibnamefont {Inden}}, \ and\ \bibinfo {author}
  {\bibfnamefont {K.}~\bibnamefont {Ishida}},\ }\href {\doibase
  10.2355/isijinternational.49.1212} {\bibfield  {journal} {\bibinfo  {journal}
  {ISIJ International}\ }\textbf {\bibinfo {volume} {49}},\ \bibinfo {pages}
  {1212} (\bibinfo {year} {2009})}\BibitemShut {NoStop}%
\bibitem [{\citenamefont {Witte}(2016)}]{Witte2016a}%
  \BibitemOpen
  \bibfield  {author} {\bibinfo {author} {\bibfnamefont {R.}~\bibnamefont
  {Witte}},\ }\emph {\bibinfo {title} {{Strain adaption in epitaxial Fe-Rh
  nanostructures}}},\ \href {http://tuprints.ulb.tu-darmstadt.de/5795/} {Ph.D.
  thesis},\ \bibinfo  {school} {Technische Universit{\"{a}}t}, \bibinfo
  {address} {Darmstadt} (\bibinfo {year} {2016})\BibitemShut {NoStop}%
\bibitem [{Note1()}]{Note1}%
  \BibitemOpen
  \bibinfo {note} {At least in the investigated alloy concentration; alloys
  with higher Fe contents may be stabilized in the cubic
  structure.}\BibitemShut {Stop}%
\bibitem [{\citenamefont {Kaspar}\ \emph {et~al.}(2014)\citenamefont {Kaspar},
  \citenamefont {Bowden}, \citenamefont {Wang}, \citenamefont {Shutthanandan},
  \citenamefont {Manandhar}, \citenamefont {van Ginhoven}, \citenamefont
  {Wirth},\ and\ \citenamefont {Kurtz}}]{Kaspar2014}%
  \BibitemOpen
  \bibfield  {author} {\bibinfo {author} {\bibfnamefont {T.~C.}\ \bibnamefont
  {Kaspar}}, \bibinfo {author} {\bibfnamefont {M.~E.}\ \bibnamefont {Bowden}},
  \bibinfo {author} {\bibfnamefont {C.}~\bibnamefont {Wang}}, \bibinfo {author}
  {\bibfnamefont {V.}~\bibnamefont {Shutthanandan}}, \bibinfo {author}
  {\bibfnamefont {S.}~\bibnamefont {Manandhar}}, \bibinfo {author}
  {\bibfnamefont {R.~M.}\ \bibnamefont {van Ginhoven}}, \bibinfo {author}
  {\bibfnamefont {B.~D.}\ \bibnamefont {Wirth}}, \ and\ \bibinfo {author}
  {\bibfnamefont {R.~J.}\ \bibnamefont {Kurtz}},\ }\href {\doibase
  10.1016/j.tsf.2013.09.067} {\bibfield  {journal} {\bibinfo  {journal} {Thin
  Solid Films}\ }\textbf {\bibinfo {volume} {550}},\ \bibinfo {pages} {1}
  (\bibinfo {year} {2014})}\BibitemShut {NoStop}%
\bibitem [{\citenamefont {Kauffmann-Weiss}\ \emph {et~al.}(2014)\citenamefont
  {Kauffmann-Weiss}, \citenamefont {Hamann}, \citenamefont {Reichel},
  \citenamefont {Siegel}, \citenamefont {Alexandrakis}, \citenamefont {Heller},
  \citenamefont {Schultz}, \citenamefont {Ludwig},\ and\ \citenamefont
  {F{\"{a}}hler}}]{Kauffmann-Weiss2014}%
  \BibitemOpen
  \bibfield  {author} {\bibinfo {author} {\bibfnamefont {S.}~\bibnamefont
  {Kauffmann-Weiss}}, \bibinfo {author} {\bibfnamefont {S.}~\bibnamefont
  {Hamann}}, \bibinfo {author} {\bibfnamefont {L.}~\bibnamefont {Reichel}},
  \bibinfo {author} {\bibfnamefont {A.}~\bibnamefont {Siegel}}, \bibinfo
  {author} {\bibfnamefont {V.}~\bibnamefont {Alexandrakis}}, \bibinfo {author}
  {\bibfnamefont {R.}~\bibnamefont {Heller}}, \bibinfo {author} {\bibfnamefont
  {L.}~\bibnamefont {Schultz}}, \bibinfo {author} {\bibfnamefont
  {A.}~\bibnamefont {Ludwig}}, \ and\ \bibinfo {author} {\bibfnamefont
  {S.}~\bibnamefont {F{\"{a}}hler}},\ }\href {\doibase 10.1063/1.4870759}
  {\bibfield  {journal} {\bibinfo  {journal} {Apl Materials}\ }\textbf
  {\bibinfo {volume} {2}},\ \bibinfo {pages} {046107} (\bibinfo {year}
  {2014})}\BibitemShut {NoStop}%
\bibitem [{\citenamefont {Franke}\ and\ \citenamefont
  {Neusch{\"{u}}tz}()}]{Franke}%
  \BibitemOpen
  \bibfield  {author} {\bibinfo {author} {\bibfnamefont {P.}~\bibnamefont
  {Franke}}\ and\ \bibinfo {author} {\bibfnamefont {D.}~\bibnamefont
  {Neusch{\"{u}}tz}},\ }in\ \href {\doibase 10.1007/10757285_82} {\emph
  {\bibinfo {booktitle} {Binary Systems. Part 4: Binary Systems from Mn-Mo to
  Y-Zr}}}\ (\bibinfo  {publisher} {Springer-Verlag},\ \bibinfo {address}
  {Berlin/Heidelberg})\ pp.\ \bibinfo {pages} {1--4}\BibitemShut {NoStop}%
\bibitem [{\citenamefont {Hao}\ and\ \citenamefont {Sholl}(2012)}]{Hao2012}%
  \BibitemOpen
  \bibfield  {author} {\bibinfo {author} {\bibfnamefont {S.}~\bibnamefont
  {Hao}}\ and\ \bibinfo {author} {\bibfnamefont {D.~S.}\ \bibnamefont
  {Sholl}},\ }\href {\doibase 10.1021/jp210573a} {\bibfield  {journal}
  {\bibinfo  {journal} {Journal of Physical Chemistry C}\ }\textbf {\bibinfo
  {volume} {116}},\ \bibinfo {pages} {2045} (\bibinfo {year}
  {2012})}\BibitemShut {NoStop}%
\bibitem [{\citenamefont {Lee}\ \emph {et~al.}(2015)\citenamefont {Lee},
  \citenamefont {Brown}, \citenamefont {Alp}, \citenamefont {Ren},
  \citenamefont {Lu}, \citenamefont {Woo},\ and\ \citenamefont
  {Johnson}}]{Lee2015}%
  \BibitemOpen
  \bibfield  {author} {\bibinfo {author} {\bibfnamefont {E.}~\bibnamefont
  {Lee}}, \bibinfo {author} {\bibfnamefont {D.~E.}\ \bibnamefont {Brown}},
  \bibinfo {author} {\bibfnamefont {E.~E.}\ \bibnamefont {Alp}}, \bibinfo
  {author} {\bibfnamefont {Y.}~\bibnamefont {Ren}}, \bibinfo {author}
  {\bibfnamefont {J.}~\bibnamefont {Lu}}, \bibinfo {author} {\bibfnamefont
  {J.-J.}\ \bibnamefont {Woo}}, \ and\ \bibinfo {author} {\bibfnamefont
  {C.~S.}\ \bibnamefont {Johnson}},\ }\href {\doibase
  10.1021/acs.chemmater.5b02918} {\bibfield  {journal} {\bibinfo  {journal}
  {Chemistry of Materials}\ }\textbf {\bibinfo {volume} {27}},\ \bibinfo
  {pages} {6755} (\bibinfo {year} {2015})}\BibitemShut {NoStop}%


\end{thebibliography}

\end{document}